# Platinum-adsorbed Defective 2D Monolayer Boron Nitride: A Promising Electrocatalyst for $O_2$ Reduction Reaction

## Lokesh Yadav[1] and Srimanta Pakhira[1, 2]*

[1] Theoretical Condensed Matter Physics and Advanced Computational Materials Science Laboratory, Department of Physics, Indian Institute of Technology Indore (IIT Indore), Simrol, Khandwa Road, Indore-453552, Madhya Pradesh, India.
[2] Theoretical Condensed Matter Physics and Advanced Computational Materials Science Laboratory, Centre for Advanced Electronics (CAE), Indian Institute of Technology Indore (IIT Indore), Simrol, Khandwa Road, Indore-453552, Madhya Pradesh, India.

*Corresponding author: spakhira@iiti.ac.in (or) spakhirafsu@gmail.com

## Abstract

The wide bandgap and strong covalent bonds of hexagonal boron nitride (hBN) had long been thought to be chemically inert. Due to its inertness with saturated robust covalent bonds, the pristine 2D monolayer hBN cannot be functionalized for applications of energy conversion. Therefore, it is necessary to make the 2D hBN chemically reactive for potential applications. Here, we have computationally designed a single nitrogen (N) and boron (B) di-vacancy of the 2D monolayer hBN, noted by $V_{BN}$ defective-BN (d-BN), to activate the chemical reactivity, which is an effective strategy to use the d-BN for potential applications especially in electrochemistry. Single Pt atom adsorbed on the defective area of the $V_{BN}$ d-BN acts as a single-atom catalyst (SAC) which exhibits distinctive performances for $O_2$ reduction reaction (ORR). First-principles based dispersion-corrected periodic hybrid Density Functional Theory (DFT-D) method has been employed to investigate the equilibrium structure and properties of the Pt-adsorbed 2D monolayer defective boron nitride (Pt-d-BN). The present study shows the semiconducting character of Pt-d-BN with an electronic bandgap of 1.30 eV, which is an essential aspect of the ORR. The ORR mechanism on the surface of 2D monolayer Pt-d-BN follows a $4e^-$ reduction route because of the low barriers to OOH formation and dissociation, $H_2O_2$ instability, and water production on the Pt-d-BN surface. Here, both the dissociative and associative ORR mechanisms have been investigated, and it is found that the associative mechanism with the ORR pathway is more thermodynamically favorable.

Therefore, it can be mentioned here that the 2D monolayer Pt-d-BN exhibits a high selectivity for the four-electron reduction pathway. According to the calculations of the relative adsorption energy of each step in ORR, the Pt-d-BN is anticipated to exhibit substantial catalytic activity. These findings are significant because they provide an additional understanding of the ORR process on the metal atom-adsorbed d-BN and a new method for producing inexpensive materials with strong electrocatalytic activity for various applications of fuel cells.

**Introduction**

Now-a-days, researchers have an emphasis on working for clean and renewable energy sources to replace traditional ones, and most of the energy sources in our society are conventional energy sources such as coal, petroleum, and compressed natural gas (CNG). One of the drawbacks of using these conventional energy sources is the emission of greenhouse gases. New methods of producing clean and green energy, high-capacity energy devices, and effective manufacturing are needed to address problems like global climate change and energy requirements of the future. Fuel cells and metal-air batteries, which are used in solar cells, electric vehicles, and other applications, are promising technologies to meet future energy requirements.[1–4] These new energy sources have a low operating temperature, high power density, high energy conversion efficiency, and nearly zero emission of greenhouse gases. An electrochemical device (such as a fuel cell) uses two redox processes to transform the chemical energy of a fuel into electrical energy. The crucial step in the fuel cells is the oxygen reduction reaction (ORR), which dramatically affects how energy is converted from chemical energy to electrical energy. Two different mechanisms, a two-electron transfer reaction and a four-electron transfer reaction mechanism that yields $H_2O_2$ and $H_2O$ as a product, respectively, can be used to perform ORR.[5] Because of the higher efficiency of the four-electron ($4e^-$) transfer reaction, it is used to promote the pathway of $O_2$ into $H_2O$ in fuel cells. The reaction pathway of the four-electron transfer mechanism is the dominant phenomenon for catalysts made up of Platinum.[4,6]

The cathodic ORR is one of the crucial energy conversion reactions in proton exchange membrane fuel cells (PEMFCs) with a significant performance since it has relatively slower kinetics than the anodic reaction.[7] It should be mentioned here that low price and remarkable catalyst performance are very important for the cathode ORR of fuel cells as the ORR process is well known to have sluggish reaction kinetics. Therefore, the development of high-performance ORR catalysts is essential and requires a better understanding of the underlying ORR mechanism. Pt-based ORR electrocatalysts are widely used to accelerate the ORR kinetics as they have a lower overpotential value.[6] Although, noble metal-free catalysts have been studied recently, especially, Pt-based catalysts are still more efficient in terms of catalytic performance and long-term durability.[8–11] The widespread commercialization of Pt-based catalysts is severely hindered by their high cost. In a recent theoretical study, a new class of ORR electrocatalysts was developed consisting of the Pt alloyed with intermediate transition metals like Sc or Y, and they could have more significant activity than the pristine Pt electrodes.[12]

New possibilities to improve the efficiencies of catalysts have opened in recent years due to the emergence of two-dimensional (2D) materials. These 2D materials have a higher value of the specific surface area and surface atomic ratio than traditional materials, which is particularly useful in the catalysis industry.[13] A large number of research groups have reported the production of 2D ultrathin Pt-containing nanosheets and monolayers demonstrating outstanding ORR catalytic activity compared to commercial catalysts.[14–17] Recent studies have shown that metal-free carbon materials, especially N-doped and B-doped carbon materials, are efficient catalysts for ORR. Due to the ability of N-species to take electrons, which results in an overall positive charge on the nearby C atom on which $O_2$ is adsorbed.[18] This N-doped carbon has improved ORR catalytic activity.[18] On the other hand, because of the B atom electron-accepting properties, there is an adsorption of $O_2$ on the B atom in the case of B-doped carbon.[19] The enhanced ORR catalytic activity was achieved on a carbon electrode which was doped with B and N atoms.[19–21]

2D monolayer hexagonal boron nitride (hBN) with a geometrical structure resembling graphene can be created by replacing all the carbon atoms in graphene with boron (B) and nitrogen (N) atoms. Recently, an N-doped hBN monolayer has been demonstrated theoretically to be capable of electrocatalytic ORR.[22,23] To use hBN-based materials as an electrocatalyst for ORR, there is a requirement for electronic communication to the hBN surface, but the pristine hBN is an insulator with a broad bandgap that ranges from 3.6 to 7.1 eV (depending on the experimental techniques).[22–25] Since hBN has a wider energy gap than graphene, it is more difficult to make it potentially applicable. Several approaches have also been employed, including functional group adsorption, metal doping and hydrogenation.[26,27] Due to the $sp^3$ bonding, we can say possibly a size mismatch between the metal atom and the B/N atom, these approaches bring the structural deformation of the pristine 2D planar geometry of the hBN.

Additionally, it is demonstrated that the conducting edge states and vacancy defects in the hBN give rise to semiconducting characteristics. Hence, the hBN sheet supported by conducting materials could function as an ORR-active electrocatalyst.

The hBN nanosheet, a structural counterpart of graphene, has also drawn great interest because of its special characteristics and vital applications.[28–30] 2D monolayer hBN sheet, for illustration, demonstrates excellent thermal stability and high electrical resistance. More precisely, the hBN sheet still maintains its stability at temperatures up to 1000 K, whereas graphene-based materials maintain its stability up to 800 K.[31] Further a theoretical research revealed that boron monovacancies have more formation energies than nitrogen monovacancies.[32] Therefore, they are more energetically favorable. The chemical reactivity of 2D hBN monolayer considerably has increased due to these defect sites and by encasing their defect sites with metal atoms.[32–34] Due to its chemical inertness behavior, the defect-free hBN sheet has minimal interaction with the deposited metal atoms. The defective BN sheet may be more helpful for creating new catalysts by metal atom doping or adsorption.[32–35]

Feng et al. reported that the surface-bound metal nanostructures, especially substrate confined single metal atom, exhibit improved $H_2$ evolution catalytic performance.[33] Very recently, Lei et al. experimentally showed that the formation of different vacancies in the d-BN could be controllably functionalized with single metal atoms by the spontaneous reduction of metal cations which are active for the $H_2$ evolution reaction.[36] They synthesized Pt-adsorbed d-BN material and computationally studied the stability and properties of both the Pt-d-BN and d-BN materials. It should be mentioned that the 2D monolayer d-BN material can reduce the metal cations.[36] Following their work, we have computationally designed a pristine 2D sheet (monolayer) of hexagonal boron nitride (hBN), a single nitrogen (N) and boron (B) di-vacancy of the 2D monolayer hBN (3x3 supercell), noted by $V_{BN}$ defective-BN (d-BN), and Pt-adsorbed 2D monolayer defective boron nitride (Pt-d-BN) material, as shown in Figure 1. It is well known that the 2D pristine hBN nanomaterial cannot act as an effective electrocatalyst for the ORR in a direct manner, and it has a large electronic band gap ($E_g$), making it an electrical insulator. It has an insufficient number of active catalytic sites, which prevents it from practical utilization towards the catalytic reactions. The pristine 2D monolayer h-BN often contains various types of defects, such as vacancies, interstitial atoms, dislocations, grain boundaries, and edges. A good number of studies have already been published for single-atom doping in the pristine 2D hBN, but, unfortunately, d-BN has not been too much explored and still, there is an opportunity to investigate the d-BN with its applications in electrochemistry. With the introduction of $V_{BN}$ di-vacancy, the 2D monolayer d-BN has become chemically active and adsorbed a single Pt atom in the defective region of it. A single Pt atom has been considered to be adsorbed around the defective area of the $V_{BN}$ d-BN followed by the experimental work[36], and we hypothesize that this single Pt atom has become efficient to reduce the $O_2$ during the ORR process and it can act as a single atom catalyst (SAC) towards ORR. In continuation of the SAC work, we have studied the ORR using Pt-d-BN material in the present manuscript which suggests further to the experimentalists to perform the experiment. The equilibrium structure of the Pt adsorbed d-BN was obtained by the first-principles based hybrid B3LYP-D3 (in short DFT-D)[36] method, and the electronic properties (i.e., the electronic band structure, position of the Fermi energy level ($E_F$), electronic band gap ($E_g$), and total density of states (DOS)) of the Pt-d-BN material were computed by employing the same DFT-D method. It was computationally found that the electronic band gap of the pristine hBN, $V_{BN}$ di-vacancy (i.e., defective boron nitride (in short d-BN)), and Pt-d-BN are 6.23 eV, 4.89 eV, and 1.30 eV, respectively, obtained by the DFT-D method. After the Pt absorption, the system raises the carrier concentration and reduces the electronic band gap from 6.23 eV to 1.30 eV, which in turn into the electronic semi-conductivity. This suggests a novel strategy to improve the ORR performance of the 2D Pt-d-BN sheet. In the present investigation, both the associative and dissociative ORR mechanisms have been studied in detail. All the reaction steps with the intermediates involved in the subject reaction have been computed along with significant changes in adsorption energy (ΔE), relative Gibbs free energy (ΔG), electronic band structure and DOS of each ORR step. The computed results indicate that the 2D monolayer Pt-d-BN material has an excellent electrocatalytic activity with a high 4e$^-$ reduction pathway selectivity where Pt acts as a single atom catalyst (SAC). Hence, we can say that the 2D monolayer Pt-d-BN material is an efficient electrocatalyst for ORR resulting it promising for fuel cells applications in energy technology. The purpose of this work is to predict the properties of the Pt-d-BN material with the possible application in fuel cells and proper strategies of elucidating the mechanisms of the 2D d-BN materials as electrocatalysts towards ORR.

## Methodology and Computational Details

A microscopic understanding of the role of a catalyst in the form of 2D monolayer slab structure, molecular and atomic regimes is now possible because of the developments in the field of computational materials science. Recent developments in Gaussian basis sets, Gaussian-types of orbitals (GTOs), pseudopotential, and supercomputers have facilitated simulations for probing the surface phenomena of diverse catalysts, analyzing various catalytic activities, and examining the chemical reaction pathways.[37] The reaction intermediates with the relative energy barriers have an important role in understanding how an atom adheres to a catalytic surface and provides information in many electrocatalytic applications of catalysts.[38] Based on the first principles methods, we have employed the periodic hybrid dispersion-corrected hybrid density functional theory (in short DFT-D) approach to investigate the chemical reaction pathway along with the mechanism of ORR taken place on the surface of 2D monolayer Pt-d-BN material. The analysis of the different properties, namely thermodynamical properties, electronic properties and also reaction mechanisms of the 2D monolayer structure, can be accomplished very well by computational techniques based on quantum mechanics.[39,40]

**(a) Theory and Computational Modeling**

First principles-based periodic dispersion-corrected hybrid density functional theory (DFT-D) technique was used to obtain the equilibrium structure of the 2D pristine hBN monolayer, d-BN and Pt-d-BN materials.[39,40] To create the 2D d-BN sheet, the equilibrium unit cell of the pristine 2D monolayer hBN was increased to a 3 × 3 supercell. Following the previous work,[36] a single nitrogen (N) and boron (B) di-vacancy of the 2D monolayer hBN (3x3 supercell), noted by $V_{BN}$ defective-BN (d-BN), has been computationally designed, and it has found that a single Pt atom has been reduced in the defective area. To put it another way, we used a hexagonally symmetric 2D hBN material to simulate a 3 × 3 supercell slab structure to build the 2D Pt-adsorbed d-BN monolayer material around the $V_{BN}$ defective area of the d-BN. The 2D monolayer pristine hBN, $V_{BN}$ defective-BN, and Pt-d-BN materials were computationally studied to determine the electronic properties, equilibrium structure, and band structure along with the total density of states (DOS) employing the same level of theory. The equilibrium lattice parameters of the pristine 2D monolayer hBN were found to be a = b = 2.499 Å with layer group symmetry "*P-6m2*". The values of equilibrium lattice constants of the Pt-d-BN were calculated to be a = 7.384 Å, b = 7.572 Å, along with layer group symmetry "*P1*". The equilibrium structures of the 2D monolayer pristine hBN, $V_{BN}$ d-BN and Pt-d-BN are illustrated in Figure 1a-c.

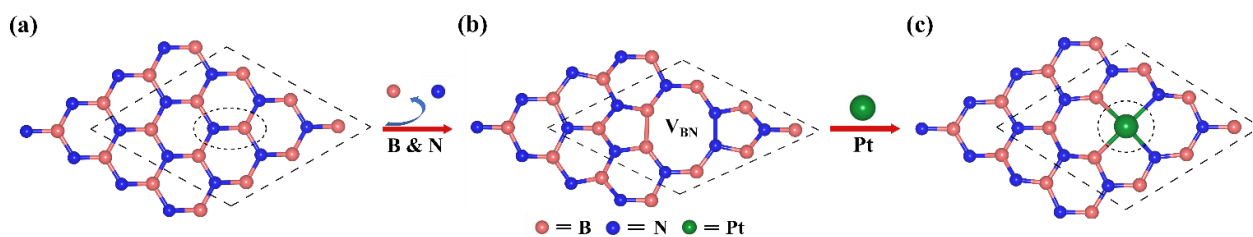

**Figure 1.** (a) The top view of the pristine 2D monolayer hBN, (b) the top view of the 2D monolayer $V_{BN}$ defective-BN ($V_{BN}$ d-BN), and (c) the top view of the 2D monolayer platinum-adsorbed $V_{BN}$ defective-BN (Pt-d-BN) materials computed by the DFT-D method are presented here. (The dotted lines with a parallelogram represent the boundary of 3 × 3 supercells.)

**(b) Periodic DFT-D Calculations**

The periodic hybrid dispersion-corrected density functional theory (DFT-D) method B3LYP-D3 (based on first principles) implemented in CRYSTAL17 suite code was used to determine the equilibrium structures and properties of the 2D monolayer pristine hBN, $V_{BN}$ defective-BN, and Pt-d-BN materials studied here.[41–44] This CRYSTAL17 software uses Gaussian types of atomic basis sets, which is more efficient than plane waves based codes for the hybrid DFT-D calculations.[45–48] In the current calculations, triple-ζ valence with polarization quality (TZVP) Gaussian basis sets were used for the O, H, B, and N atoms, and for the Pt atom, we used the Pt doll 2004 basis set with the relativistic effective core potentials (ECPs). It is important to consider the weak van der Waals (vdW) dispersion effects during the calculations because of the presence of vdW interactions between the atoms in the layers of the hBN, $V_{BN}$ defective-BN, and Pt-d-BN nanosheets, as well as among all the atoms. The semi-empirical Grimme's 3rd order (-D3) corrections of dispersion have been used in this study to account for all the long-range vdW interactions. In addition to provide the reliable and attractive geometry of the monolayer 2D structure of the systems considered here, DFT-D (i.e., B3LYP-D3) is an extremely useful method because of the least effect of spin contamination on the energy and density in the current calculations.[49–52] In each calculation, the electronic self-consistency scale was set to $10^{-7}$ a.u. Two-dimensional (2D) vacuum slabs were created for these materials so that the electrostatic potential could be considered in these computations, and the energies were reported in reference to the vacuum. The height of the vacuum cell was set to 500 Å (the 2D slab model in the CRYSTAL17 algorithm with no periodicity in the z-direction).[41–44] Hence, the present computations took the vacuum region nearly of the order of 500 Å. Although this strategy is different from plane-wave codes such as VASP and Quantum Espresso but both lead to the same conclusions.

Now, equilibrium geometries, structures, and electronic properties (such as the electronic band structures, Fermi energy ($E_F$), electronic band gap ($E_g$), and the total density of states (DOS)) of the 2D monolayer structure of pristine hBN, $V_{BN}$ defective-BN, and Pt-d-BN were calculated at the same level of theory. All integrations of the first Brillouin zone were sampled on 4×4×1 Monkhorst-Pack k-mesh grids for all the systems considered here. A threshold of $10^{-7}$ a.u. was selected as the evaluation point for the convergence of energy, forces, and electron density. For all the materials (i.e., the pristine hBN, d-BN, and Pt-d-BN), the highly symmetric ***Γ-M-K-Γ*** route in the first Brillouin zone was selected as the k-vector path for visualizing the band structures. A total number of eight electronic energy bands (number of four valence bands and conduction bands) has been drawn around the Fermi energy level ($E_F$). The total DOS was computed using the atomic orbitals of the elements B, N, O, H, and Pt. The present work used electrostatic potential calculations, i.e., the energy of both the band structures and DOS is computed with respect to the vacuum. A visualization code (VESTA)[53] was used to graphically represent and analyze all the optimized 2D layer structures.

**(c) O$_2$ Reduction Reaction Mechanism**

Nørskov and coworkers developed a method for the analysis of several elementary stages, including various protonation states to characterize the O$_2$ reduction reaction pathway and they proposed the possible reaction pathway.[54] The present DFT-D study reveals that the reduced Pt in the defective area of the $V_{BN}$ defective-BN ($V_{BN}$ d-BN) surface is the chemically and energetically favorable site for O$_2$ adsorption, which is the initial step of the ORR. The sequences of the O$_2$ reduction reaction have been explained by both the 2e$^-$ and 4e$^-$ transfer mechanisms. The ORR mechanism is completed by obtaining two different reaction steps, which may contain associative or dissociative reaction pathways.[55] The associative and dissociative reaction steps are represented by Eq. (1) and (2), respectively.

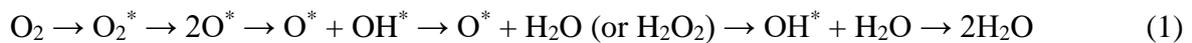

$$O_2 \rightarrow O_2^* \rightarrow 2O^* \rightarrow O^* + OH^* \rightarrow O^* + H_2O \text{ (or } H_2O_2) \rightarrow OH^* + H_2O \rightarrow 2H_2O \quad (1)$$

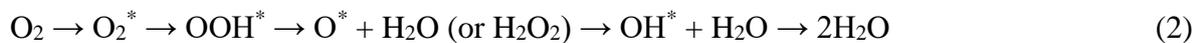

$$O_2 \rightarrow O_2^* \rightarrow OOH^* \rightarrow O^* + H_2O \text{ (or } H_2O_2) \rightarrow OH^* + H_2O \rightarrow 2H_2O \quad (2)$$

where * represents an active site of the catalysts, and these active sites adsorb the reaction intermediate species throughout the reactions. In the above two equations, the adsorption of protons (H$^+$) and electrons (e$^-$) has been ignored. During the ORR process, these 4e$^-$ and 2e$^-$ transfer pathways result in the formation of H$_2$O and H$_2$O$_2$ as the final product, respectively. We have focused here on both the dissociative and associative mechanisms because it has been found that both the mechanisms are thermodynamically and energetically favorable for the 2D monolayer Pt-d-BN material.

**(D) Equations for Energy Calculations and Thermodynamic Analysis**

For the 2D monolayer Pt-d-BN, we have calculated the catalytic performance by evaluating the adsorption energy of the adsorbate at the catalytic surface at each reaction step during the ORR process. Also, the binding energy of protons ($H^+$) and electrons ($e^-$) has been included in this calculation. The following equation has been used to calculate the adsorption energy ($E_{ads}$) of the species that are present during the ORR process on the adsorption system (Pt-d-BN).

$$E_{ads} = E_{total} - (E_{adsorbent} + E_{adsorbate})$$

Where $E_{total}$, $E_{adsorbent}$, and $E_{adsorbate}$ are the total energies of the adsorbed system, adsorbent (isolated Pt-d-BN sheet), and adsorbate species, respectively, in the gas phase. A negative value of $E_{ads}$ suggests that the process is exothermic, and it is occurring because the adsorbed oxygenated intermediates are attached to the thermodynamically stable surface of the 2D monolayer Pt-d-BN. This material has been considered in the present study to explore all the reaction steps of ORR with respect to computational hydrogen electrode model (CHE) or standard hydrogen electrode (SHE).[2,4,54,55] It should be noted here that CHE model is a widely-accepted approach in the realm of electrochemical reactions and it predominantly addresses the thermodynamics of the system.[2,4,54,55] We have also calculated the changes of free energy ($\Delta G$) of all the ORR steps on the surface of 2D monolayer Pt-d-BN, determined by $\Delta G = \Delta E + \Delta ZPE - T\Delta S + \Delta G_{pH}$. Where $\Delta E$ is the total energy change directly obtain from DFT calculations, $\Delta ZPE$ and $\Delta S$ are the zero-point energy difference and the entropy difference between the adsorbed state and the gas phase respectively and T is the temperature of the system. $\Delta G_{pH}$ is the free energy contributions due to variations in $H^+$ concentration, and here we have taken its value zero for the acidic medium. It should be mentioned here that the thermodynamic analysis and Gibbs free energy calculations during the ORR have been performed at room temperature (T = 298.15 K). In this work, the slab structure of 2D monolayer Pt-d-BN is used as the standard reference state for all adsorption energy calculations.

## Results and Discussions

At first, we studied the equilibrium structure and electronic properties of the pristine 2D monolayer structure of the hBN sheet or slab. The 2D equilibrium structure of the hBN is depicted in Figures 1a and 2a. The unit cell was expanded three times (3x3 supercell) along the x and y directions, and the z-direction remained unchanged, i.e., we created a 3x3 supercell structure of the pristine 2D monolayer hBN. The B3LYP-D3 method was used to obtain the equilibrium structure and band structure along with the total DOS of the pristine 2D monolayer hBN, as shown in Figures 2a-c. The results of our present DFT-D investigation reveal that the pristine 2D hBN sheet exhibits hexagonal *P-6m$_2$* symmetry with lattice constants a = b = 2.449 Å, α = β = 90°, and γ = 120º. The equilibrium B-N bond length is about 1.442 Å, and the interatomic distance between the nearest neighbors, N-N and B-B, is found to be about 2.49 Å. The Fermi energy level ($E_F$) of the pristine 2D hBN sheet was calculated to be at -7.20 eV and is represented by a dotted blue line in Figure 2b-c. Calculations of the band structure of the pristine 2D monolayer hBN consist of four conduction bands and four valence bands, which have been computed and plotted around the Fermi energy level depicted in Figure 2b and 2c. It has a direct electronic bandgap ($E_g$) of about 6.23 eV, which demonstrates the insulating properties of the pristine hBN sheet. The total electron density of states (DOS) calculations has been performed at the same level of theory as depicted in Figure 2c, demonstrating that the electronic band gap of pristine 2D hBN sheet is around 6.23 eV which is consistent with obtained band from the band structures calculations, and there is almost negligible electron density around the Fermi level. Due to this large band gap, the pristine 2D hBN sheet behaves as an insulator. Hence, it can conclude that the 2D monolayer hBN cannot be employed as an electrocatalyst to improve the kinetics of ORR in fuel cells.

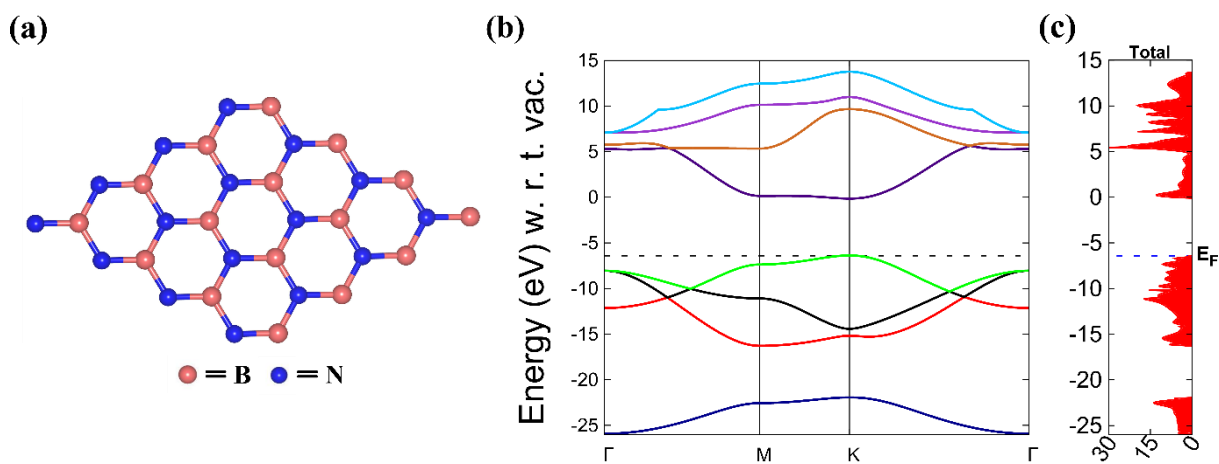

**Figure 2.** The equilibrium structure and electronic properties of the 2D monolayer pristine hBN sheet: (a) top view, (b) electronic band structure, and (c) total density of states (DOS) are displayed.

It has been found that the pristine 2D monolayer hBN cannot function as effective catalyst for the ORR due to its large electronic bandgap and inertness.[36] The performance of the 2D hBN can be dramatically improved by creating defects, and the defective boron nitride shows a dependence on the location and distribution of defects, which can reduce the metal ions or can adsorb the metal atoms and reduce the metal ions. Here, we have introduced a single nitrogen (N) and boron (B) di-vacancy of the 2D monolayer hBN (3x3 supercell), noted by $V_{BN}$ defective-BN (d-BN), and studied the structural and electronic properties of this material. The 2D equilibrium structure and band structures along with the total DOS of the $V_{BN}$ d-BN, are obtained by the B3LYP-D3 method and depicted in Figure 3a-c. The results of our present DFT-D investigation reveal that after introducing $V_{BN}$ defects into the 2D pristine hBN sheet, the lattice constants were increased, and it was found that the new lattice constants are a = b = 7.401 Å, α = β = 90°, and γ = 126.87° with the *P1* symmetry. The average B-N bond length around the defective area is about 1.459 Å, and the equilibrium B-B and N-N bond lengths are about 1.730 Å and 1.569 Å, respectively, at the boundaries of the defective area obtained by the same level of theory. The electronic band structures of the $V_{BN}$ d-BN have been computed along the highly symmetric *Γ-M-K-Γ* k-direction in the first Brillouin zone following the pristine 2D h-BN for comparison. The Fermi energy level ($E_F$) of the 2D monolayer $V_{BN}$ d-BN is found at -6.10 eV, which is represented by a dotted blue line in Figure 3b-c. It should be noted here that the position of the $E_F$ has been shifted by an amount of 1.10 eV towards the conduction bands, as displayed in the electronic band structures and total DOS calculations displayed in Figure 3b-c.

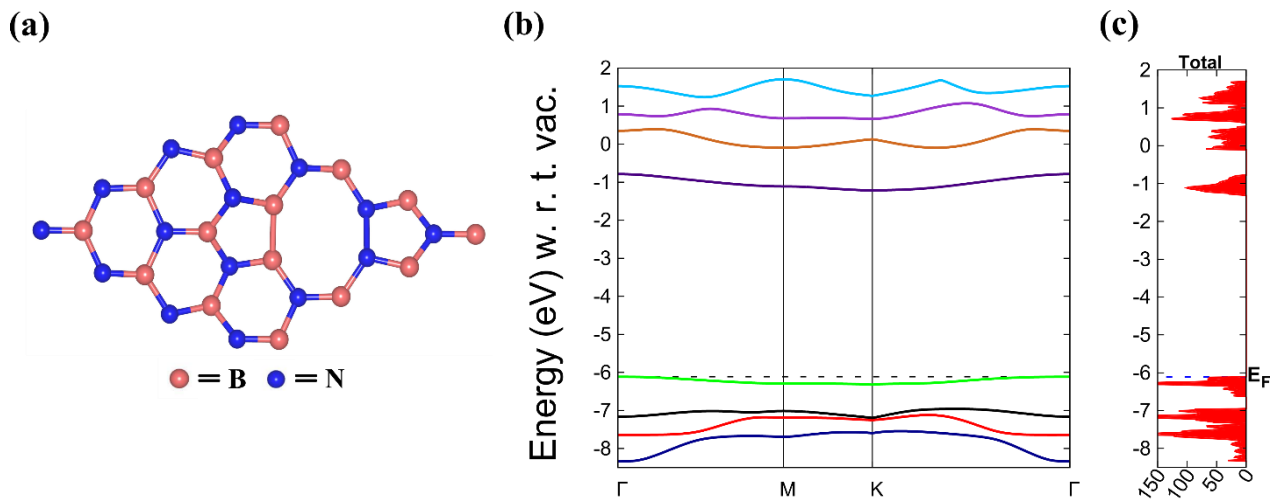

**Figure 3.** The equilibrium structure of 2D monolayer $V_{BN}$ d-BN material: (a) top view, (b) electronic band structure, and (c) total density of states (DOS) are displayed.

The present study shows that the value of electronic band gap ($E_g$) of the $V_{BN}$ d-BN sheet has been reduced by an amount of 1.34 eV compared to the pristine 2D hBN, and the electronic band gap of it is about 4.89 eV, as depicted in Figure 3b, obtained by the DFT-D method. The DOS shows the same value of band gap energy and a small value of electron density just below the Fermi energy level, as depicted in Figure 3c, resulting in a high band gap semi-conductor. Moreover, it was computationally found that the absorption of a Pt atom at the BN-divacancy site of d-BN is energetically preferable compared to the monovacancy site. Therefore, we first investigated the equilibrium structure, geometry, and stability of this Pt-adsorbed BN-divacancy i.e., 2D monolayer Pt-d-BN. The equilibrium structure reveals that the platinum atom has slightly shifted toward the center of the defective region ($V_{BN}$) and formed four covalent bonds with its nearby B and N atoms, which is shown in Figure 1c. As the size of the Pt atom is large relative to both the B and N atoms, the adsorbed Pt atom shifts upward from the plane of the 2D d-BN sheet, which is displayed in Figure 4b. The average equilibrium Pt-N and Pt-B bond lengths are 2.04 Å and 2.08 Å, respectively, which is less than the total atomic radius of the participating atoms.

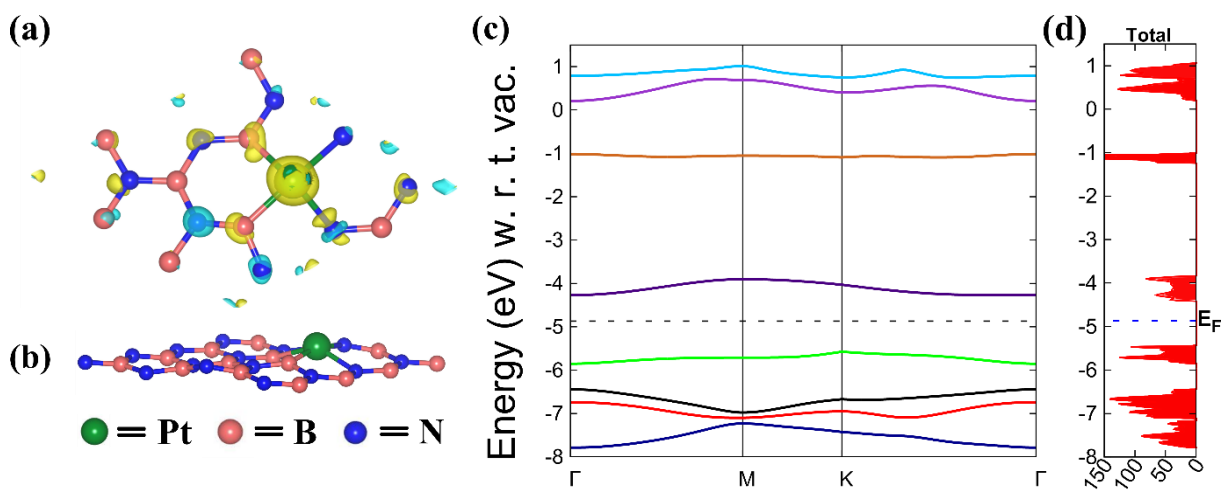

**Figure 4.** (a) The electronic spin densities, (b) side view, (c) electronic band structure, and (d) total density of states of the equilibrium structure of 2D monolayer Pt-d-BN are displayed. The yellow color represents the positive part of the wave function that corresponds to the up-spin (α-spin) of the electrons, and the sky-blue color represents the negative part of the wave function that corresponds to the down-spin (β-spin) of the electrons in the 2D monolayer Pt-d-BN material.

We have used the DFT-D technique to study the electron spin density along with an equilibrium 2D monolayer structure of the 2D Pt-d-BN material (3x3 supercell). Figure 4a-b represents electronic spin density and a side view of the equilibrium structure of the 2D Pt-d-BN sheet. Electron spin density calculations help us to locate the possible sites of unpaired electrons, i.e., electron cloud around the defective area in the Pt-d-BN material and give us an idea about the active site for a reaction to happen. Then, we calculated the electronic properties of this material, including its band structure and total DOS, at the equilibrium geometry, which is shown in Figures 4c-d. After the absorption of Pt into the 2D monolayer $V_{BN}$ d-BN, the values of the lattice constants increased. The new calculated lattice constants are found to be a = 7.384 Å, b = 7.572 Å, α = β = 90˚, and γ = 117.70˚ with *P1* symmetry obtained by the same DFT-D method. The Fermi energy level ($E_F$) is at -5.55 eV, which is moved towards the conduction bands as depicted in Figure 4c, and it has been represented by the dotted blue line in Figure 4c. It should be mentioned here that the electronic band gap ($E_g$), which was initially around 6.23 eV in the case of 2D pristine hBN monolayer sheet, has now been decreased to 1.3 eV after Pt absorption. As a result, the 2D monolayer Pt-d-BN has become a semiconductor with a small band gap around 1.30 eV, as displayed in Figures 4c-d. The lower value of the electronic bandgap of the 2D Pt-d-BN materials increases the electron concentration, and it improves the electronic semi-conductivity in nature, and it may help to facilitate higher electron transfer towards the reactants during the ORR process. This provides a novel strategy to make the Pt-adsorbed 2D monolayer d-BN, i.e., Pt-d-BN, for the effective performance of ORR. Therefore, the single Pt atom adsorbed in the d-BN can act as an efficient single-atom electrocatalyst (SAC) for ORR. Thus, an enhanced electrocatalytic ORR activity is anticipated from the 2D monolayer Pt-d-BN material, which may be useful for fuel cells. The equilibrium lattice constants (a and b), various average bond lengths, symmetry, and electronic band gap of the 2D monolayer hBN, d-BN, and Pt-d-BN materials are summarized in Table 1.

**Table 1** Equilibrium structural and lattice parameters (in Å) of the pristine hBN, $V_{BN}$ d-BN and Pt-d-BN materials obtained by the DFT-D method are summarized here.

| System | Lattice parameters (Å) | 2D Layer group and symmetry | Electronic band gap ($E_g$) | Fermi energy ($E_F$) | Average bond distance between atoms (Å) | | |
|---|---|---|---|---|---|---|---|
| | | | | | B-N | Pt-B | Pt-N |
| hBN | a = b = 2.449 | ***78, P-6m2*** | 6.23 | -7.20 | 1.442 | - | - |
| $V_{BN}$ d-BN | a = b = 7.401 | ***1, P1*** | 4.89 | -6.10 | 1.459 | - | - |
| Pt-d-BN | a = 7.384, b = 7.572 | ***1, P1*** | 1.30 | -5.55 | 1.440 | 2.089 | 2.047 |

To evaluate the stability of 2D monolayer Pt-d-BN material, the binding energy of Pt atom in the 2D monolayer Pt-d-BN material was calculated as follows: $E_b = E_{total}(Pt\text{-}d\text{-}BN) - E_{total}(Pt) - E_{total}(V_{BN}\_d\text{-}BN)$, where $E_{total}$ is the total energy corresponding to the different species mentioned in the brackets. The calculated value of binding energy of Pt atom in the 2D monolayer Pt-d-BN material is about 0.10 eV. The large $E_b$ of Pt atoms on this $V_{BN}\_d$-BN also makes the diffusion of the adsorbed Pt atom to its neighboring hollow site which is considerably difficult in terms of the large endothermicity and high energy barrier, thus vigorously excluding the clustering of adsorbed Pt atom on this substrate. These results indicate that Pt atom can be stably adsorbed on the defective area of BN sheet both thermodynamically and kinetically, thus ensuring the stability of Pt-d-BN sheet. It should be mentioned that the 2D monolayer d-BN material can reduce the metal cations as Lei et al.[36] experimentally showed that due to the vacancy induced free radicals and Fermi level shifts, d-BN can be controllably functionalized with single metal atoms by the spontaneous reduction of metal cations; mono-metallic or bi-metallic clusters can also be effectively reduced. In essence, these vacancies are the reactive sites to reduce metal cations spontaneously on the d-BN surface, as well as the PL emission sites within the visible region. A single Pt atom can also be directly bonded on BN at the vacancy site which has been discussed in the present investigation. There might be possible to form clusters of Pt atoms in the d-BN, but, followed by previous work, we have performed only Pt stably anchored in the 2D monolayer d-BN as the single atom which is our main focus of the present investigation. While electrocatalysis is a proof-of-concept application, the experimental findings[36] insights into the selective reduction of metal ions on activated BN are universal to create stabilized single metal atoms (like Pt), that can be used for other emerging applications, e.g., other catalysts, such as ORR, HER, etc.[36]

To study the thermodynamical stability of the materials, we have simulated the Raman spectrum, Raman active frequencies and amplitudes of vibrational modes using the B3LYP-D3 method, which is implemented in the ab initio-based CRYSTAL17 suite code. A set of 4×4×1 Monkhorst-Pack k-point grids were used in these calculations. The standard approach to calculate the vibrational spectra is to calculate harmonic vibrational normal modes. In other words, a harmonic vibrational analysis was performed at the equilibrium structure of the 2D monolayer, pristine h-BN, $V_{BN}\_d$-BN and Pt-d-BN materials to obtain the Raman spectra. This static method is built upon the assumption that the potential energy surface at an energy minimum can be approximated by a harmonic potential. Figure 5 shows the simulated Raman spectrum for a) hBN, b) V$_{BN}$_d-BN and c) Pt-d-BN materials. We have remarked the active Raman spectra of hBN, and it shows two peaks at the wavenumber 877.77 cm$^{-1}$ and 1372.35 cm$^{-1}$. On the other hand, in the Raman spectrum of V$_{BN}$_d-BN, the vibration of d-BN with a wavenumber 1339.93 cm$^{-1}$ ($\upsilon_1$) is the most intense peak (taken as 100%). The other vibrations are also observed at both left and right side of this intense peak, but these have less intensity as compared to $\upsilon_1$ vibrational mode. All other vibrations are nearly (10 - 40%) of this intense peak. In the case of 2D monolayer Pt-d-BN material, three intense peaks are found in our present computational study. One vibrational mode ($\upsilon_1$) is at wavenumber 449.39 cm$^{-1}$ (taken as 100%), the second one ($\upsilon_2$) is at wavenumber 1374.83 cm$^{-1}$ which is nearly 90% intense as compared to the highest Raman active vibrational mode $\upsilon_1$, and third one ($\upsilon_3$) is at wavenumber 1141.03 cm$^{-1}$ which is nearly 60% intense as compared to $\upsilon_1$. The Raman active spectrum in this case is spread from 400 - 1500 cm$^{-1}$. All other vibrations are less intense nearly less than 50% of the intensity of $\upsilon_1$.

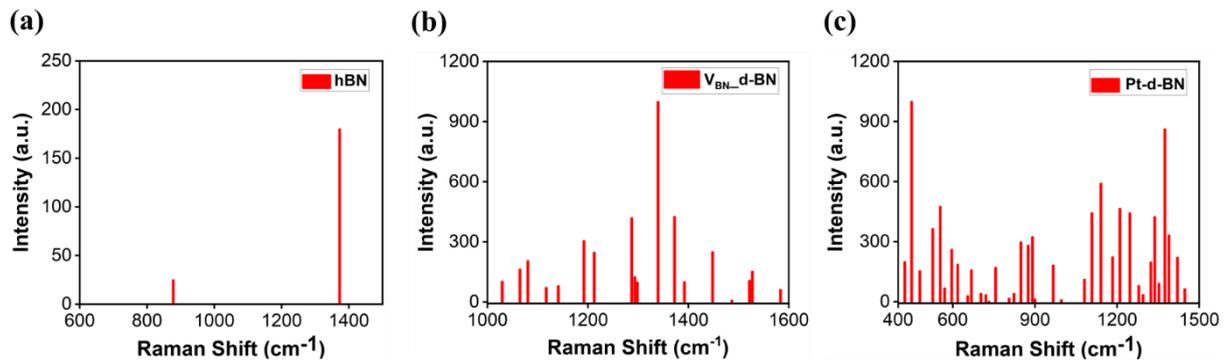

**Figure 5.** Raman spectra of the 2D monolayer a) hBN, b) V$_{BN}$_d-BN and c) Pt-d-BN materials are shown here computed by the B3LYP-D3 method.

After obtaining the equilibrium structure with the electronic properties of the 2D monolayer Pt-d-BN, we investigated the electrocatalytic activity of the Pt-d-BN towards the ORR. This study aimed to investigate the potential of this 2D Pt-d-BN nanosheet as an efficient ORR electrode material for fuel cells. In the present investigation, we have explored all the ORR pathways by using the periodic slab structure of the 2D Pt-d-BN material based on the computational hydrogen electrode (CHE). The ORR consists of 4 electrons (e$^-$) and 4 protons (H$^+$) transfer during the reaction of each O$_2$ molecule, as well as other reaction intermediates such as O$^*$ (oxygen), OH$^*$ (hydroxyl groups), and OOH$^*$ (super hydroxyl groups) that have been adsorbed onto the active sites of the 2D Pt-d-BN material during the completion of ORR.

Therefore, it can be mentioned here that this material exhibits a high four-electron reduction pathway selectivity for effective ORR.

Before looking into the possible mechanisms for ORR on the surface of the 2D monolayer Pt-d-BN material, it is essential to note that the reduction of adsorbed $O_2$ involves the following two reaction steps: First, by the reaction $O_2^* \rightarrow 2O^*$, it can immediately dissociate into two active atomic oxygen (O) which is known as dissociative reaction mechanism; second, it may be hydrogenated to form the $OOH^*$ species by the reaction $O_2^* + H^+ + e^- \rightarrow OOH^*$ which is known as associative reaction mechanism. In the first reaction, the O-O bond gets broken, resulting in the adsorption of two individual atomic oxygens on the surface of the 2D Pt-d-BN material. One of the O atoms from the $O_2$ after dissociation, remains at the Pt atom, while another O atom creates a bond with the nearest N atom and forms the N-O bond. In contrast to the first step, the second one involves the addition of the $H^+ + e^-$ to the $O_2^*$, which was adsorbed by the Pt-d-BN, resulting in a yield of $OOH^*$ on the surface of the Pt-d-BN. Moreover, the $OOH^*$ species cannot easily dissociate into $O^* + OH^*$ on the Pt-d-BN surface, unlike the $O_2^*$ molecule, which may be the cause of the high activation energy values. We are quite curious whether this novel 2D structure of the Pt-d-BN material is a promising candidate for fuel cell applications, as other Pt-based nanomaterials have displayed outstanding performance for launching the sluggish ORR.

In the present work, we have included both the associative and dissociative mechanisms for ORR and describe each possible reaction pathway on the surface of the 2D monolayer Pt-d-BN material. According to the dissociative mechanism, the $O_2$ dissociates on the active site of the Pt-d-BN (i.e., Pt area), in which, one $O^*$ creates a weak bond with the Pt atom while other moves to favorable neighboring N. After the subsequent steps, it makes a bond with two pairs of electrons ($e^-$) and protons ($H^+$) during the formation of intermediates which results in the formation of $H_2O$ molecule. In the associative mechanism, during the adsorption of $O_2$ on the active site of the Pt-d-BN surface, the O-O bond does not break in the subsequent steps. Instead of that it can attract H ($H^+ + e^-$), resulting in the formation of $OOH^*$, which is further reduced to form $H_2O$. The entire reaction pathways and all possible reaction intermediates are shown in Figures 6a-b, respectively, for both the dissociative and associative reaction mechanisms of the ORR process on the surface of the 2D monolayer Pt-d-BN.

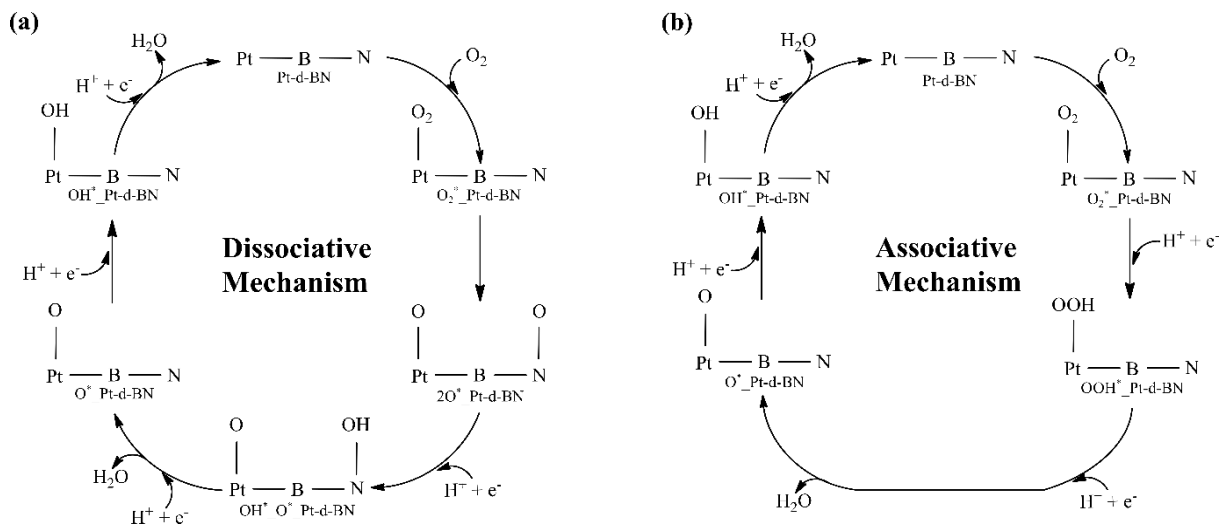

**Figure 6.** Schematic representation of the $O_2$ reduction reaction pathways on the surfaces of 2D monolayer Pt-d-BN for (a) dissociative mechanism and (b) associative mechanism.

It is a challenging task to do computational modeling of the $O_2$ reduction reaction intermediates because of the need to incorporate the effects of solvent/surface on the adsorbed intermediates, the high electric field in the double layer, the free energy of solvated reactants and the free energy of electrons in the solid as a function of potential. A simple method to solve this issue is to link the general trend to gas phase modeling or solid surface reaction modeling. The results of DFT computation are accurate for gas species and solid surface reactions, and it is simple to compute the electrolytic processes at the gas-solid interface. Here, we have used a small segment of this Pt-d-BN material with one oxygen molecule adsorbed as a model for the ORR pathways to explain the reaction mechanism.

**Dissociative ORR mechanism:**

After calculating the electronic properties and confirming the electronic semi-conductivity of the 2D monolayer Pt-d-BN, we were interested in learning more about the ORR activity of the given material as an electrode, more specifically, a cathode of a fuel cell. This may show excellent performance for launching the sluggish ORR in fuel cells. We have studied each of the reaction intermediates following the ORR mechanism depicted in Figure 6a, which illustrates the proposed four-electron (4e$^-$) transfer reaction pathway or four-electron transfer mechanism, and it is represented as $O_2 + 4H^+ + 4e^- \rightarrow 2H_2O$. It is believed that the adsorption character at the active site of the catalyst is the most crucial parameter of an efficient ORR.

# Dissociative Path

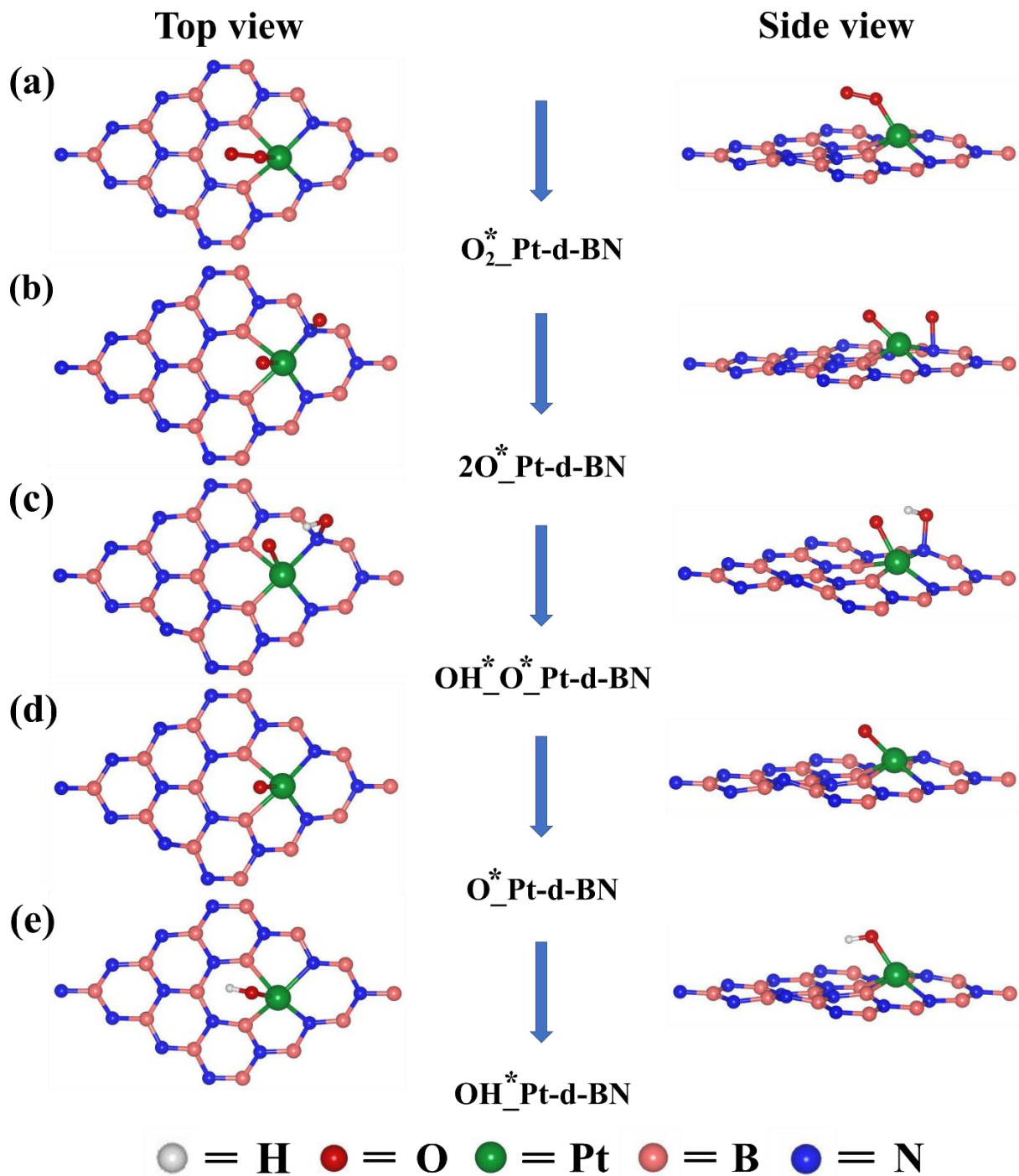

**Figure 7.** Top view and side view of various equilibrium reaction intermediates during the ORR process followed by dissociative mechanisms on the surface of Pt-d-BN sheet: (a) $O_2^*$_Pt-d-BN, (b) $2O^*$_Pt-d-BN, (c) $OH^*$_$O^*$_Pt-d-BN, (d) $O^*$_Pt-d-BN, (e) $OH^*$_Pt-d-BN.

We have computationally explored the $O_2$ reduction reaction steps with the intermediates and ORR mechanism on the surface of the 2D monolayer Pt-d-BN material followed by relative free energy and adsorption energy calculations. The initial step of the ORR begins with the adsorption of oxygen molecule ($O_2^*$) on the active surface of the 2D monolayer structure of the Pt-d-BN at the Pt-site (which was considered the active site of the catalyst) shown in Figure 7a by considering the computational model system with computational hydrogen electrode (CHE) conditions. When the $O_2^*$ has reached the surface of 2D monolayer Pt-d-BN material, the O−O bond length is significantly stretched from 1.230 Å to 1.311 Å computed by the DFT-D method, which indicates that the oxygen molecule has been effectively activated towards the subject reaction. Following the adsorption of $O_2^*$ onto the surfaces of the Pt-d-BN, it is found that the equilibrium Pt-O and O-O bond lengths are 1.935 Å and 1.311 Å, respectively, obtained by the DFT-D method. We have also calculated that the change of Gibb's free energy ($\Delta G$) during this stage is about -0.66 eV and the relative adsorption energy ($\Delta E$) is about -0.84 eV, which means that it is energetically favorable, and the intermediate formed by the $O_2^*$ adsorption is noted by $O_2^*$_Pt-d-BN as depicted in Figure 7a. Both side view and top view of the equilibrium structures of the $O_2^*$_Pt-d-BN reaction intermediate is depicted in Figure 7a. Now, the adsorbed $O_2^*$ dissociates into $2O^*$ in the subsequent stage of the ORR process, with one oxygen atom being adsorbed at the N site (which is close to the Pt site) on the surfaces of the 2D Pt-d-BN, as noted by $2O^*$_Pt-d-BN. The top view, along with its side view of the equilibrium structure of this reaction intermediate $2O^*$_Pt-d-BN, is illustrated in Figure 7b. The equilibrium Pt-O bond length has been reduced, and it is about 1.834 Å, and the newly formed N-O bond length was found to be around 1.509 Å. This process has a shift in free energy of about +0.66 eV and in adsorption energy it is +0.68 eV, as represented by step II in the energy diagram of the dissociative ORR pathway (or potential energy surface (PES)) displayed in Figure 8a-b. The fact is that the change in free energy of this step is slightly positive, which means that it is an endothermic process. In the next step, out of the two present active sites, the electron ($e^-$) and proton ($H^+$) were simultaneously added to one of the activated oxygen atoms at the N-site, and the value of free energy change in this reaction step is -0.64 eV and adsorption energy change is -1.47 eV. This intermediate reaction step is noted by $OH^*$_$O^*$_Pt-d-BN, and the equilibrium structure is shown in Figure 7c. The results of the current investigation show that the equilibrium bond lengths of the Pt-O and N-OH are about 1.941 Å and 1.514 Å, respectively, calculated the DFT-D method. Additionally, the equilibrium bond length of O-H was calculated to be about 1.01 Å. Again, the process of addition of another electron ($e^-$) and proton ($H^+$) was repeated, and the first $H_2O$ molecule was released from the surface of the 2D Pt-d-BN electrocatalyst and resulted in a change in free energy about -2.59 eV and in adsorption energy about -1.21 eV, as shown in Figure 8 a-b. Now, only one active oxygen atom remains on the catalyst surface as shown in the proposed reaction pathway, and intermediate reaction step is noted by O$^*$_Pt-d-BN which is illustrated in Figure 7d. Additionally, in this process, the formation of H$_2$O$_2$ is a highly endothermic process. Therefore, it is out of scope in our current study of dissociative mechanisms. It is also observed that the equilibrium Pt-O bond length gets reduced, and the value of the Pt-O bond distance is about 1.834 Å calculated by the DFT-D method. Again, the addition of a single proton (H$^+$) and a single electron (e$^-$), which are coming from the anode side, resulted in the formation of the OH$^*$_Pt-d-BN intermediate, as both the proton and electron are migrated towards the remaining active site O$^*$ of the O$^*$_Pt-d-BN. This results in a change in free energy of about -1.55 eV and in adsorption energy change about -1.90 eV computed by the present DFT-D method as depicted in the energy diagram of the dissociative ORR pathway in Figure 8 a-b. Now, the equilibrium Pt-O bond length of this intermediate state OH$^*$_Pt-d-BN was slightly increased by an amount of 0.074 Å and found to be about 1.974 Å. Both the top and side views of the optimized structure are illustrated in Figure 7e. Furthermore, during the final stage of the ORR process, the second H$_2$O molecule was released after the addition of proton and electron, resulting in a change of free energy about -0.33 eV and change of adsorption energy is -0.45 eV, displayed in Figure 8 a-b. The 4e$^-$ transfer mechanism is involved in this entire ORR process, and the process of protonation of OH$^*$ to H$_2$O is a potential-limiting step. The free energy change and adsorption energy change for each reaction step (except step 2) is negative, as reported in Table 3.

    The 2D monolayer Pt-d-BN also shows high selectivity for the 4e$^-$ reduction pathway. By keeping the basal plane active, each of the reaction steps takes place at the Pt-site of the 2D monolayer of the Pt-d-BN material which indicates that Pt atom acts as a single-atom catalyst in the system. Table 2 provides a comparison of the average bond lengths, space group symmetries, and lattice constants for each reaction intermediate of the ORR. The equilibrium geometries (top view and side view) of all the reaction intermediate states involved in the ORR mechanism are depicted in Figure 7. Relative free energy, relative adsorption energy, change in adsorption energy (ΔE), and change in free energy (ΔG) of all reaction steps are displayed in the energy diagram of the dissociative ORR pathway in Figure 8a-b. It should be mentioned here that the relative adsorption energy and relative free energy of each ORR step with the change of adsorption energy (ΔE) and free energy (ΔG) of the reaction pathway are presented in Table 3.

**Table 2** Equilibrium structural parameters, lattice constants, and electronic band gap ($E_g$) parameters of various systems of the ORR steps taken place on the surface of the 2D monolayer Pt-d-BN material.

| System | Lattice parameters (Å) | Space group & symmetry | Electronic band gap ($E_g$ in eV) | Fermi energy ($E_F$) | Average bond distance (Å) | | |
|---|---|---|---|---|---|---|---|
| | | | | | Pt-B | Pt-N | Pt-O |
| $O_2^*$_Pt-d-BN | a = 7.572, b = 7.384 | *P1* | 2.097 | -6.08 | 2.126 | 2.095 | 1.935 |
| $2O^*$_Pt-d-BN | a = 7.572, b = 7.384 | *P1* | 2.615 | -6.36 | 2.156 | 2.067 | 1.834 |
| $OH^*$_$O^*$_Pt-d-BN | a = 7.572, b = 7.384 | *P1* | 4.163 | -5.67 | 2.048 | 2.091 | 1.941 |
| $O^*$_Pt-d-BN | a = 7.572, b = 7.384 | *P1* | 2.398 | -5.91 | 2.156 | 2.090 | 1.834 |
| $OH^*$_Pt-d-BN | a = 7.572, b = 7.384 | *P1* | 3.488 | -5.56 | 2.112 | 2.098 | 1.974 |
| $OOH^*$_Pt-d-BN | a = 7.572, b = 7.384 | *P1* | 3.344 | -5.40 | 2.118 | 2.096 | 1.974 |

**Table 3** Change in adsorption energy (ΔE), change in free energy (ΔG), relative adsorption energy, and relative free energy of all ORR reaction steps during the dissociative mechanism performed on the 2D monolayer Pt-d-BN material.

| ORR Steps | ΔE (eV) | Relative adsorption energy (eV) | ΔG (eV) | Relative free energy (eV) |
|---|---|---|---|---|
| Pt-d-BN → $O_2^*$_Pt-d-BN | -0.84 | -0.84 | -0.66 | -0.66 |
| $O_2^*$_Pt-d-BN → $2O^*$_Pt-d-BN | 0.68 | -0.16 | 0.66 | 0.00 |
| $2O^*$_Pt-d-BN → $OH^*$_$O^*$_Pt-d-BN | -1.47 | -1.63 | -0.64 | -0.64 |
| $OH^*$_$O^*$_Pt-d-BN → $O^*$_Pt-d-BN | -1.21 | -2.84 | -2.59 | -3.23 |
| $O^*$_Pt-d-BN → $OH^*$_Pt-d-BN | -1.90 | -4.74 | -1.55 | -4.78 |
| $OH^*$_Pt-d-BN → Pt-d-BN | -0.45 | -5.19 | -0.33 | -5.10 |

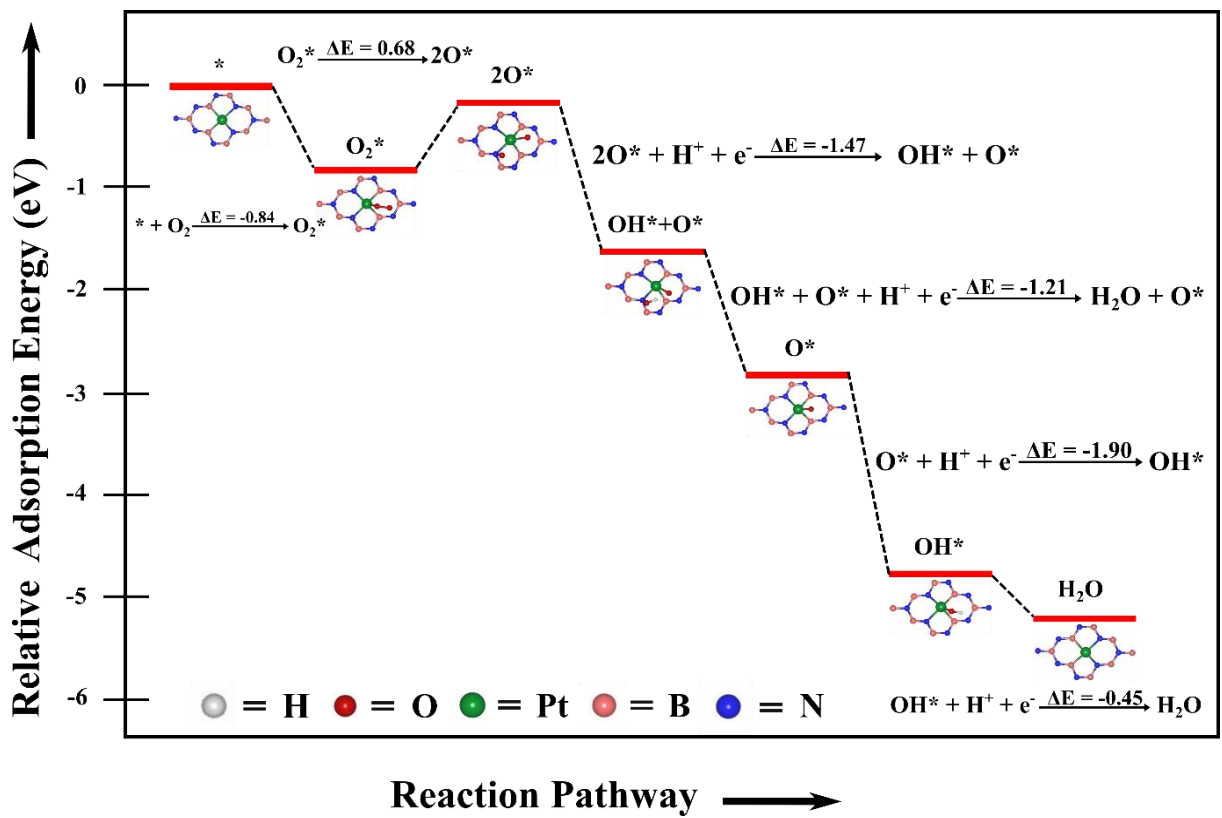

**Figure 8(a).** Relative adsorption energy diagram for the dissociative ORR pathway (energy diagram of the dissociative ORR pathway) on the surface of the Pt-d-BN material with the optimized geometries of the reaction intermediates and reaction step.

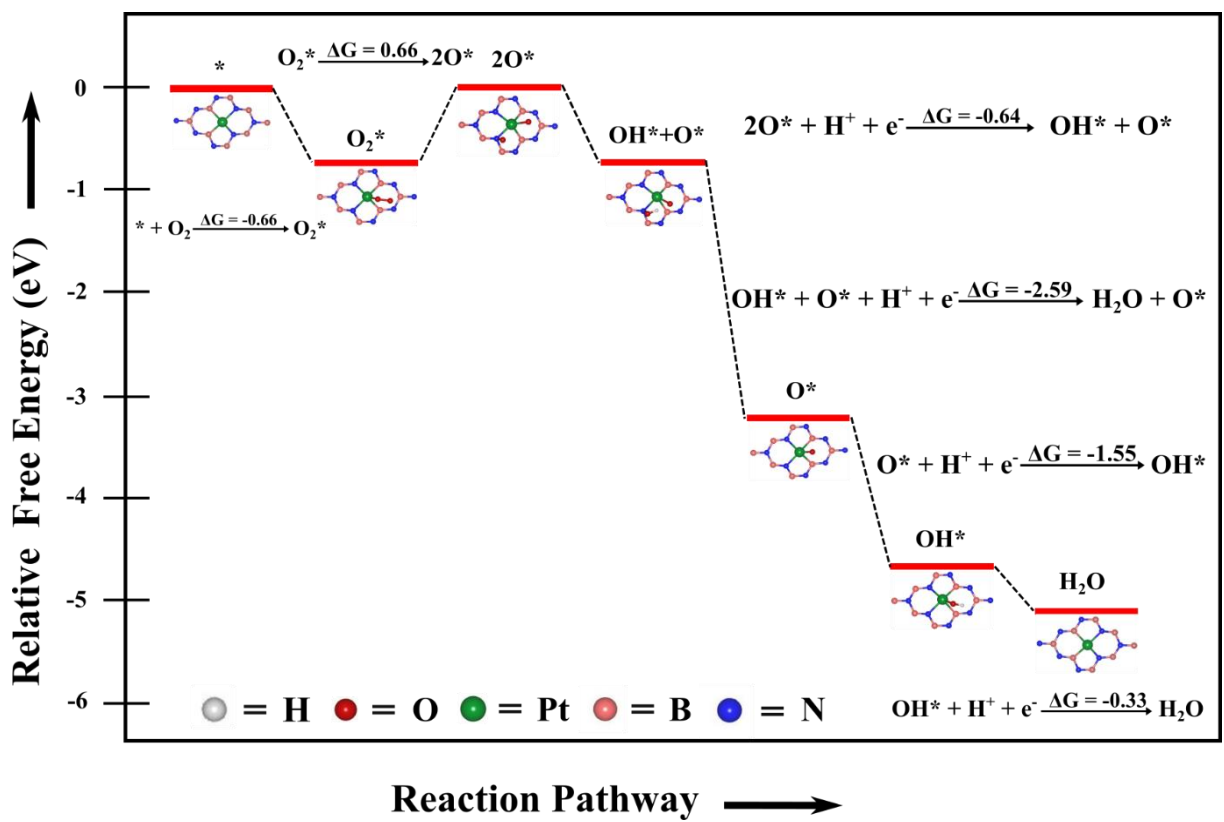

**Figure 8(b).** Relative free energy diagram for the dissociative ORR pathway (energy diagram of the dissociative ORR pathway) on the surface of the Pt-d-BN material with the optimized geometries of the reaction intermediates and reaction step.

In the present study, we have investigated the electron spin densities of these various intermediate states formed during the ORR process, as shown in Figure 9. The electron spin densities were assessed by using the same DFT-D method. The spin density estimates the differences between spin up (α) and spin down (β). In other words, this is another way to estimate the presence of unpaired electrons. Under the condition when spin up (α) and spin down (β) are not equal, there is spin polarization which results in the spin density and an unequal distribution of the electron is observed in the intermediate states. The spin density functional theory (SDFT) is the required generalization of the DFT in the presence of a magnetic field, besides the usual scalar external potential due to the nuclei. The spin density s and the electron density ρ are the two space functions acting as basic variables in SDFT. The interactions that are taking place among the electron spins within a system are also reflected by the spin density. Therefore, it plays a fundamental role for understanding magnetic phenomena of the material.[56] In the following figures, the highlighted yellow color represents the positive component of the wave function, i.e., α−spin electrons, and the sky-blue color represents the negative component of the wave function, i.e., β−spin electrons. The detailed study of the spin density distribution in a molecule is essential for interpreting spin polarization propagation in molecular complexes and crystals. It allows one to highlight and rationalize the various magnetic interactions as a function of molecular orientation and packing. The aim of the present work is to unveil the information hidden in the spin density distribution of electrons in our system. The difference in α and β electron densities results in non-zero spin polarization and electron density.

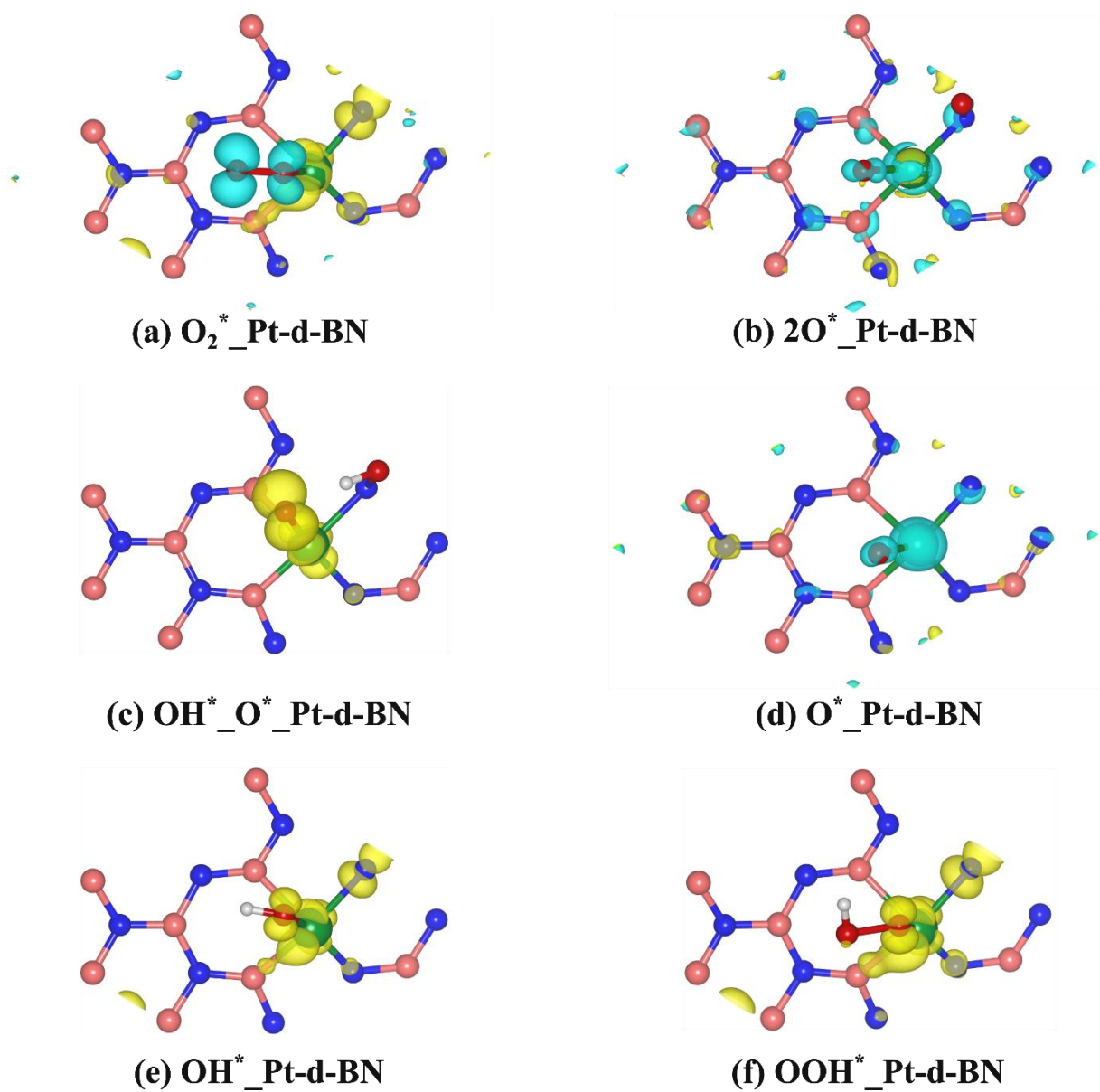

**Figure 9.** Electron spin density of the various equilibrium reaction intermediates during the ORR process: (a) $O_2^*$_Pt-d-BN, (b) $2O^*$_Pt-d-BN, (c) $OH^*$_$O^*$_Pt-d-BN, (d) $O^*$_Pt-d-BN, (e) $OH^*$_Pt-d-BN, (f) $OOH^*$_Pt-d-BN. The yellow color represents the positive part of the wave function that corresponds to the up-spin (noted by α−spin) of the electrons, and the sky-blue color represents the negative part of the wave function that corresponds to the down-spin (β−spin) of the electrons.

**Discussions of various steps of the ORR mechanism in detail:**

**Step I:** In the present work, the DFT-D method has been employed to determine the equilibrium geometry of the 2D monolayer Pt-d-BN material with the $O_2$ adsorbed Pt-d-BN, i.e., $O_2^*$_Pt-d-BN shown in Figure 7a. The present computational study reveals that the equilibrium Pt-B, Pt-N, and Pt-O bond lengths of the $O_2^*$_Pt-d-BN intermediate are 2.126 Å, 2.095 Å, and 1.935 Å, respectively, reported in Table 2. The symmetry of the $O_2^*$_Pt-d-BN reaction intermediate is **P1**, and the values of lattice constants and interfacial angles are a = 7.572 Å, b = 7.384 Å, α = β = 90° and γ = 117.70° computed by the DFT-D method. The change in free energy (ΔG) is found to be about -0.66 eV and the adsorption energy change (ΔE) is -0.84 eV, showing an exothermic behavior, which is thermodynamically favorable. The electronic band structure along with the total DOS of the $O_2^*$_Pt-d-BN intermediate reaction, are depicted in Figure 10a. The Fermi energy level ($E_F$) of the $O_2^*$_Pt d-BN is found at -6.08 eV, which is represented by a dotted blue line in Figure 10a. The value of $E_F$ has been increased by an amount of +0.5 eV after adsorbing the $O_2$ in the 2D Pt-d-BN, as depicted in Figure 10a. The lowest energy level of the conduction band was obtained at -5.12 eV, whereas the highest energy level of the valence band was calculated at -7.22 eV, both at the K point, confirming that this reaction intermediate $O_2^*$_Pt-d-BN has a direct band gap ($E_g$) and the value of $E_g$ is about 2.09 eV, as illustrated in Figure 10a. According to the calculation of the total density of states (DOS) of the $O_2^*$_Pt-d-BN, it is observed that the electrons occupied energy states both above and below the Fermi energy level, but the Fermi level itself is free from electron density, which is shown in Figure 10a.

**Step II:** The equilibrium structure of $2O^*$_Pt-d-BN intermediate is displayed in Figure 7b, in which a single oxygen atom from the $O_2$ molecule of the Pt site has been moved to the N site, then, $O_2^*$ has been dissociated on the surface of the Pt-d-BN. The values of lattice constants of this reaction intermediate $2O^*$_Pt-d-BN are a = 7.572 Å, b = 7.384 Å, and γ = 117.70° with the *P1* symmetry. This intermediate reaction step $2O^*$_Pt-d-BN (where a N-O bond has been created during this step) is in an unstable state, and the corresponding change in free energy is positive (ΔG = + 0.66 eV). It indicates that this intermediate reaction step is endothermic in nature. The same DFT-D approach has been used to determine the electronic band structures and total DOS of this reaction intermediate $2O^*$_Pt-d-BN. The Fermi energy level ($E_F$) is found to be reduced by an amount of 0.28 eV. The Fermi energy level is at -6.36 eV, shown in the electronic energy band structures calculations, and the bands have been drawn along the high directional symmetry *Γ- M – K – Γ* with respect to the vacuum depicted in Figure 10b. The Fermi energy level is represented by a dotted blue line in Figure 10b. The lowest energy level of the conduction band was calculated about -5.12 eV, whereas the highest energy level of the valence band was calculated about -7.73 eV, both at the K point, confirming that the reaction intermediate $2O^*$_Pt-d-BN has a direct band gap about 2.61 eV, which is demonstrated in Figure 10b.

**Step III:** During the ORR mechanism, the equilibrium 2D monolayer structure of the intermediate reaction step OH*_O*_Pt-d-BN has been formed. One oxygen atom has a bond with the Pt site, while the second oxygen atom has a bond with the N site, as shown in Figure 7c. The values of lattice constants corresponding to this intermediate state are a = 7.57 Å, b = 7.38 Å, and α = β = 90.0° and γ = 117.70°, with *P1* symmetry calculated by the DFT-D method. The change in Gibbs free energy (ΔG) is calculated to be about -0.64 eV and the change in adsorption energy (ΔE) is -1.47 eV which is displayed in Figure 8 a-b. After absorbing simultaneously, a single proton ($H^+$) and a single electron ($e^-$), the oxygen atom attached to the N site forms an O-H bond to get its stable configuration with an equilibrium O-H bond length of about 1.010 Å. The electronic energy band gap ($E_g$) of the intermediate reaction state OH*_O*_Pt-d-BN further was increased, and it is about 4.16 eV. The Fermi energy level has been shifted to -5.67 eV, which is highlighted by a dotted blue line in both the electronic band structure and total DOS calculations as depicted in Figure 10c. The highest energy level of the valence band (VB), in this case, is approximately -7.27 eV, while the lowest energy level of the conduction band (CB) is approximately -3.11 eV, suggesting that the intermediate OH*_O*_Pt-d-BN has a direct band gap about 4.16 eV and electron densities are not observed around the Fermi level.

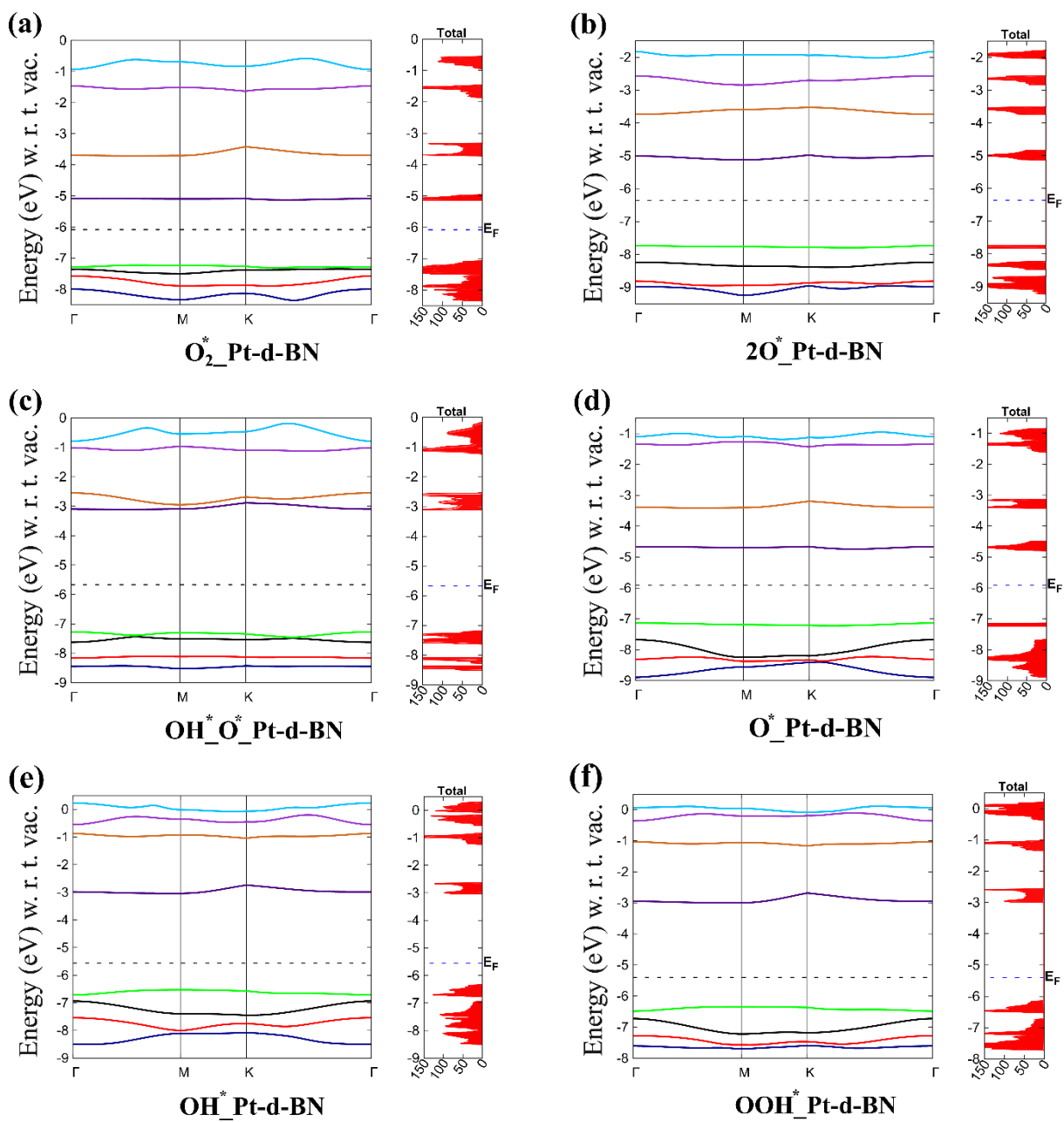

**Figure 10.** Electronic band structure and total density of states of each reaction intermediate step during ORR on the surface of 2D Pt-d-BN sheet: (a) $O_2^*$_Pt-d-BN, (b) $2O^*$_Pt-d-BN, (c) $OH^*$_$O^*$_Pt-d-BN, (d) $O^*$_Pt-d-BN, (e) $OH^*$_Pt-d-BN, and (f) $OOH^*$_Pt-d-BN.

**Step IV.** An equilibrium structure of the intermediate reaction step $O^*$_Pt-d-BN is illustrated in Figure 7d, where the OH* group attached to the N site released one H$_2$O molecule leaving one oxygen atom bound to the Pt site. The same DFT-D approach has been used to determine the band structures along with the total DOS of the reaction intermediate $O^*$_Pt-d-BN. The equilibrium Pt-O bond length in this intermediate state is about 1.83 Å. The values of equilibrium lattice constants of the $O^*$_Pt-d-BN intermediate state is found to be a = 7.57 Å, b

**27**

= 7.38 Å, α = β = 90°, and γ = 117.70° with a ***P1*** symmetry. The change in free energy is found to be about -2.59 eV and the change of adsorption energy is -1.21 eV, which is depicted in Figure 8 a-b, and it is again an exothermic process. It is computationally found that the electronic energy band gap has been reduced to approximately 2.39 eV, and the same can be observed from the band structure and total DOS calculations as depicted in Figure 10d. The Fermi energy level ($E_F$) of the reaction intermediate $O^*$_Pt d-BN is observed to be reduced by the value of 0.24 eV, and it is calculated to be about -5.91 eV which is represented by a dotted blue line in Figure 10d. In this calculation, of the electronic band structures, the lowest peak of the conduction band was calculated at about -4.74 eV, and the highest peak of the valence band was calculated about -7.13 eV, shown in Figure 10d.

**Step V.** The equilibrium structure of the last reaction intermediates $OH^*$_Pt-d-BN is formed after absorbing simultaneously a single proton ($H^+$) and a single electron ($e^-$), and the equilibrium structure is illustrated in Figure 7e. The equilibrium O-H and Pt-O bond lengths of the $OH^*$_Pt-d-BN are about 0.974 Å and 1.974 Å, respectively, computed by the DFT-D method. The values of equilibrium lattice constants were calculated to be a = 7.57 Å, b = 7.38 Å, α = β = 90°, and γ = 117.70° having a P1 symmetry. The change in free energy is calculated to be about -1.55 eV and the change in adsorption energy is -1.90 eV which is displayed in Figure 8 a-b. The Fermi energy level ($E_F$) of the reaction intermediate $OH^*$_Pt-d-BN is enhanced by an amount of 0.35 eV and calculated to be about -5.56 eV which is represented by a dotted blue line in Figure 10e. The lowest energy level of the conduction band was calculated about -3.03 eV, whereas the highest energy level of the valence band was calculated about -6.52 eV, confirming that the intermediated $OH^*$_Pt-d-BN has a direct band gap about 3.48 eV.

**Associative Mechanism of the ORR**

We have also studied the associative mechanism of the ORR on the surface of the 2D monolayer Pt-d-BN material. The illustration of the associative mechanism of ORR is demonstrated in Figure 6b. In the very first step of the associative mechanism, the adsorption of the $O_2$ molecule occurred on the Pt site of the 2D monolayer Pt-d-BN which is similar to the first step of the dissociative ORR mechanism. In other words, the $O_2$ adsorption on the surfaces of the 2D monolayer Pt-d-BN material is also the 1$^{st}$ reaction step in the case of associative ORR process i.e., the first step of both the associative and dissociative ORR mechanisms are the same which is nothing but $O_2$ activation by adsorbing it on the active surfaces of the electrocatalysts. All the structural parameters and electronic properties of this 2D equilibrium intermediate state $O_2^*$_Pt-d-BN have already been discussed in the dissociative part. The equilibrium structure (both the top and side views) with the reaction intermediated during the ORR process followed the associative reaction mechanism is illustrated in Figure 11a-d, and the detailed reaction pathway with the adsorption energy of this initial step of the associative mechanism (i.e., energy diagram of the associative ORR pathway) is displayed in Figure 12. In the next step, the adsorbed $O_2^*$ on the Pt site captures one proton ($H^+$) and one electron ($e^-$), resulting in the formation of the $OOH^*$_Pt-d-BN intermediate, which differs from the dissociative mechanism. The equilibrium Pt-B and Pt-N bond lengths are about 2.118 Å and 2.096 Å, respectively, computed by the DFT-D method. The slight increment in Pt-O bond length of this intermediate step was observed and it is about 1.974 Å, and the newly formed O-H bond length is about 0.977 Å. Both the top and side views of the equilibrium structure of the $OOH^*$_Pt-d-BN intermediate are illustrated in Figure 11b. The free energy change of this intermediate state is about -0.85 eV with the relative adsorption energy ($\Delta E$)about -1.16 eV which is shown in energy diagram of the associative ORR pathway in Figure 12 a-b. The negative values of both the change of free energy and change of the adsorption energy during the associative mechanism of ORR process taken place on the surfaces of the Pt-d-BN show that the reaction is exothermic and thermodynamically favorable in nature. In contrast to the dissociative mechanism where both the values of free energy change and the adsorption energy change of the second step ($O_2^* \rightarrow 2O^*$) are positive, which means the reaction is endothermic. However, the associative mechanism results in a decrease in free energy, which indicates the spontaneous formation of $OOH^*$_Pt-d-BN under favorable thermodynamic conditions from $O_2^*$_Pt-d-BN. The equilibrium structures of the reaction steps corresponding to the associative ORR mechanism are illustrated in Figure 11.

The equilibrium lattice constants of the reaction intermediate $OOH^*$_Pt-d-BN are a = 7.572 Å, b = 7.384 Å, and $\gamma$= 117.70° with ***P1*** symmetry. The equilibrium structure, electronic band structures and the total DOS of this intermediate reaction step $OOH^*$_Pt-d-BN are computed by using the same DFT-D method as shown in Figure 10f. The position of the Fermi energy level is at -5.40 eV which is represented by a dotted blue line in Figure 10f. The $E_F$ level has been shifted towards the conduction bands by an amount of 0.68 eV compared to the intermediate $O_2^*$_Pt-d-BN. It also has a direct band gap of 3.34 eV.

# Associative Path

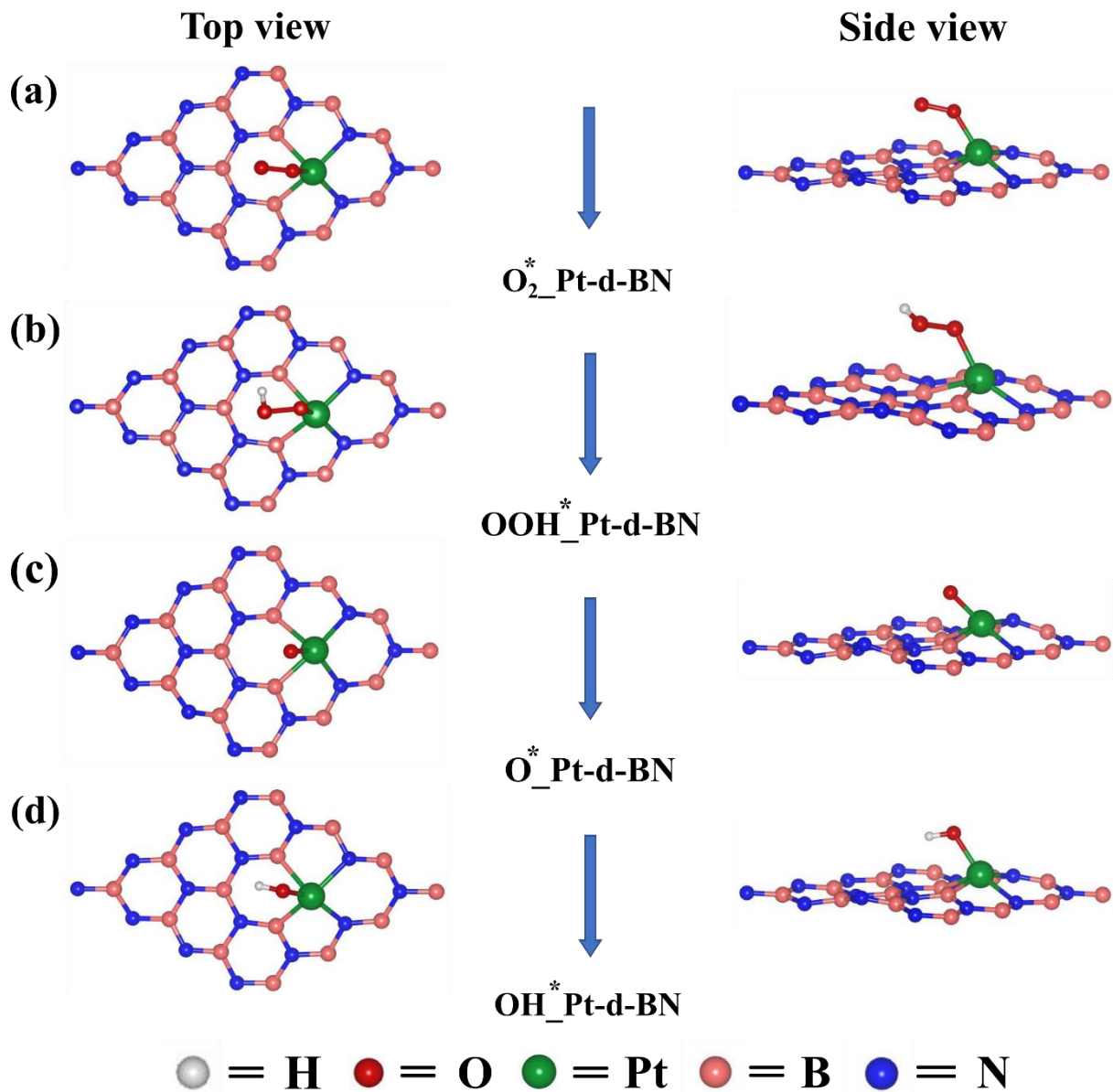

**Figure 11.** Top view and side view of various equilibrium reaction intermediate during the ORR process followed by associative mechanisms on the surface of Pt-d-BN sheet: (a) $O_2^*$_Pt-d-BN, (b) $OOH^*$_Pt-d-BN, (c) $O^*$_Pt-d-BN, and (d) $OH^*$_Pt-d-BN.

After the formation of the reaction intermediate $OOH^*$_Pt-d-BN, again, an electron and proton are simultaneously added to this system ($OOH^*$_Pt-d-BN). Now, there are two possible pathways in this step. The first possibility is the removal of the $H_2O$ molecule followed by the 4e$^-$ pathway, which results in a change in free energy of -1.72 eV and adsorption energy change of -0.83 eV. The second possibility is the removal of the $H_2O_2$ molecule followed by the 2e$^-$ pathway, which results in a change in free energy of -0.02 eV (adsorption energy change is -0.48 eV). Here, the first possibility is more favorable because of its larger value of negative adsorption energy. In short, we can say that the two-electron pathway could be significantly suppressed by the four-electron pathway under normal working conditions. As a result of this step, $O^*$_Pt-d-BN reaction intermediate is formed, and the equilibrium structure of this system is shown in Figure 11c. Additionally, the system $O^*$_Pt-d-BN undergoes the process of protonation, which is followed by the addition of a single electron and a single proton, resulting in the formation of $OH^*$_Pt-d-BN. The calculated values of change of free energy and adsorption energy are about -1.55 eV and -1.90 eV, respectively, and the reaction step is both favorable and feasible in the thermodynamic regime. The equilibrium structure of the intermediate reaction state $OH^*$_Pt-d-BN is shown in Figure 11d. The detailed discussion of all the above intermediate steps is already mentioned in the dissociative mechanism section of this manuscript. The band structure and total DOS of this associative mechanism are reported in Figure 10f, and optimized equilibrium structures of all intermediate states during the associative ORR mechanism, along with their respective reaction pathways, are presented in Figure 11.

Now, again the subsequent step involves the further addition of a single electron and a single proton to proceed further ORR. Throughout the entire process, four electrons transfer mechanism is prominent, and then the $H_2O$ is eliminated from the surface of the 2D monolayer Pt-d-BN. During the elimination of $H_2O$ from the surface of this 2D monolayer Pt-d-BN material, the change of free energy in this step is about -0.33 eV and adsorption energy change is -0.45 eV computed by the DFT-D method. The change in adsorption energies ($\Delta E$), change in free energy ($\Delta G$), relative adsorption energy, and relative free energy of all the reaction steps corresponding to the associative reaction mechanism of ORR are listed in Table 4. The values of $\Delta E$, $\Delta G$, relative adsorption energy, and relative free energy are found to be negative which indicate that all the above-mentioned reaction steps, shown in Figure 12 a-b, are thermodynamically favorable.

**Table 4** Change in adsorption energy ($\Delta E$), change in free energy ($\Delta G$), relative adsorption energy, and relative free energy of all reaction steps of ORR during the associative mechanism performed on the 2D monolayer Pt-d-BN sheet.

| ORR Steps | $\Delta E$ (eV) | Relative adsorption energy (eV) | $\Delta G$ (eV) | Relative free energy (eV) |
| --- | --- | --- | --- | --- |
| Pt-d-BN → O$_2$*_Pt-d-BN | -0.84 | -0.84 | -0.66 | -0.66 |
| O$_2$*_Pt-d-BN → OOH*_Pt-d-BN | -1.16 | -2.00 | -0.85 | -1.51 |
| OOH*_Pt-d-BN → O*_Pt-d-BN | -0.83 | -2.83 | -1.72 | -3.23 |
| O*_Pt-d-BN → OH*_Pt-d-BN | -1.90 | -4.73 | -1.55 | -4.78 |
| OH*_Pt-d-BN → Pt-d-BN | -0.45 | -5.18 | -0.33 | -5.10 |

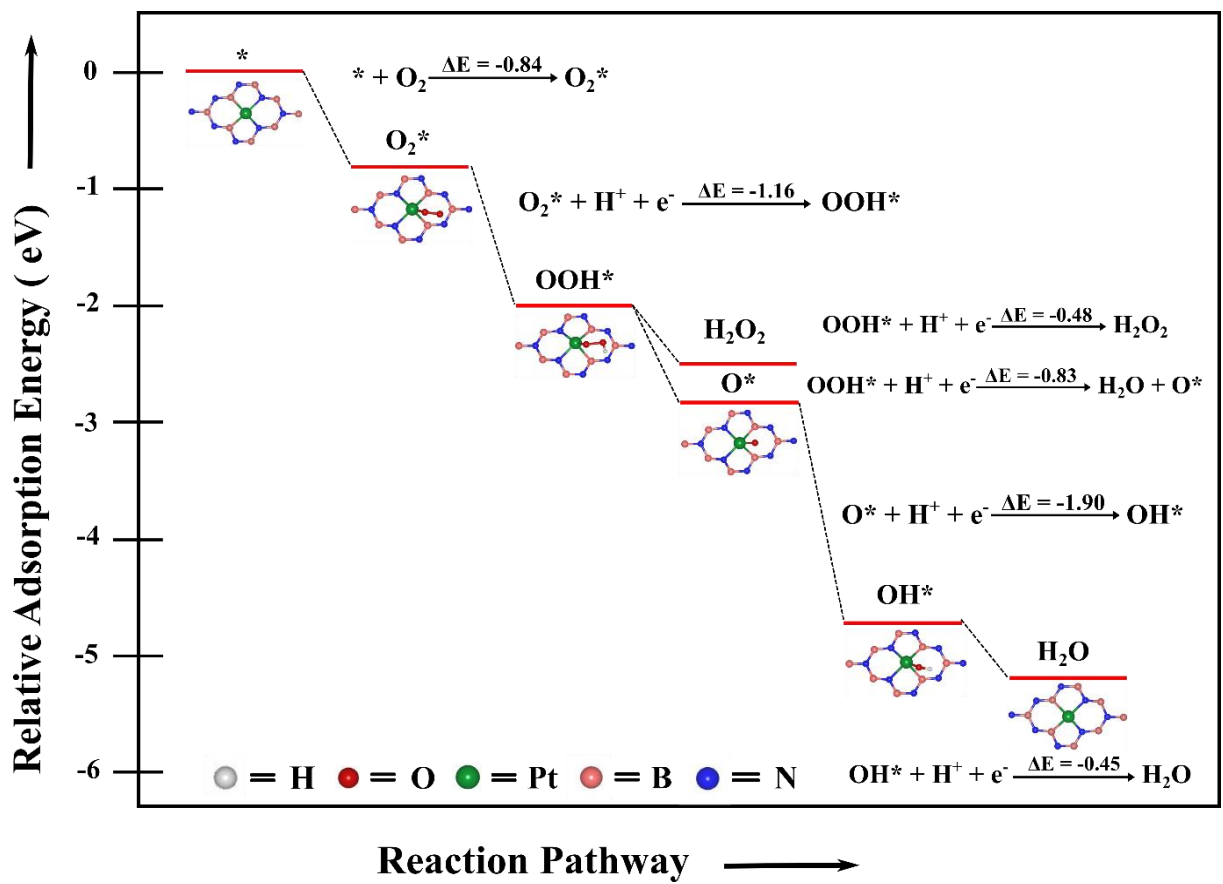

**Figure 12(a).** Relative adsorption energy diagrams for the associative ORR pathway (i.e., energy diagram of the associative ORR pathway) on the surface of the Pt-d-BN sheet with optimized geometries and reaction step.

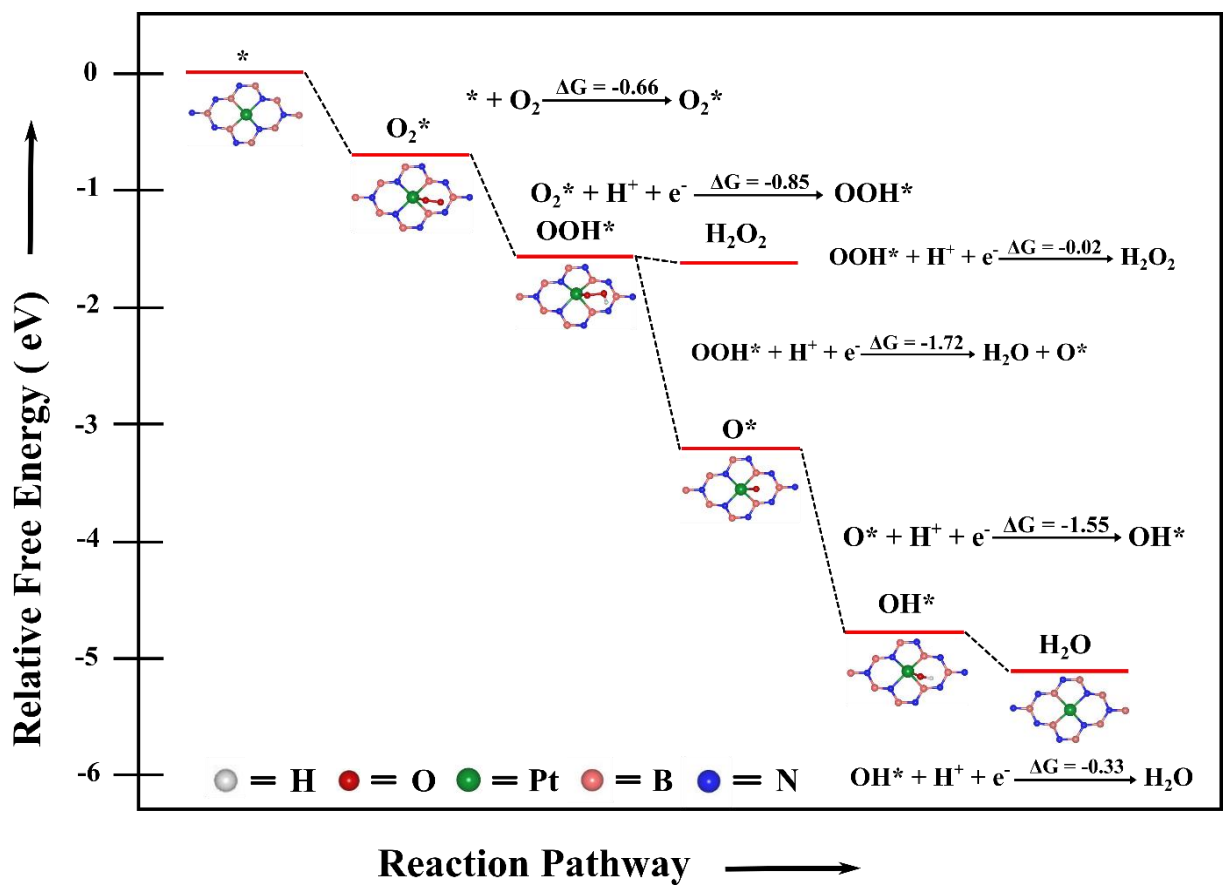

**Figure 12(b).** Relative free energy diagrams for the associative ORR pathway (i.e., energy diagram of the associative ORR pathway) on the surface of the Pt-d-BN sheet with optimized geometries and reaction step.

## Conclusions

In summary, we have computationally explored both the structural and electronic properties of the pristine 2D monolayer hBN, d-BN and Pt-d-BN materials by employing hybrid periodic DFT-D method. We have introduced a di-vacancy ($V_{BN}$) defect in the pristine 2D monolayer hBN and found that $V_{BN}$ d-BN is chemically active. A single Pt atom has been adsorbed around the defective area of the $V_{BN}$ d-BN, and the electronic band gap of the 2D monolayer Pt-d-BN has been reduced to 1.34 eV, which indicates that the 2D monolayer Pt-adsorbed d-BN is a semiconductor. In our present study, we have adopted the 2D monolayer Pt-d-BN to assist computationally in the groundwork of an efficient electrocatalyst containing a single Pt atom (which was adsorbed by the d-BN) site for highly effective electrochemical $O_2$ reduction. In short, the electrocatalytic properties of the 2D novel Pt-d-BN material have been systematically investigated on the basis of extensive DFT-D computations and the present study shows that a single Pt atom serves or performances as a SAC towards ORR. Both the associative and dissociative mechanisms during the ORR process have been explored by using the CHE model-based computations. Our present investigation has demonstrated that the 2D monolayer Pt-d-BN material is a promising ORR catalyst with excellent electrocatalytic activity and a high 4e$^-$ reduction pathway selectivity, where Pt acts as a single atom catalyst. It has found that the ORR on the surface of 2D monolayer Pt-d-BN favors the associative $O_2$ reduction reaction mechanism over the dissociative mechanism as for all the reaction pathways, free energies are negative, and all the steps of potential energy curves are downhill in the associative mechanism. Our current research indicates that the ORR might be carried out effectively with substantially improved reaction kinetics by using the Pt-d-BN 2D monolayer, and it could be a very good candidate to substitute Pt-based electrodes in fuel cells. Although it is a challenging electrode material to be used for practical purposes in fuel cells and battery technologies, but this study suggests that it could be possible to develop a 2D monolayer Pt-d-BN material in the laboratory and utilize it as an efficient ORR catalyst in the coming years for commercial applications. We also expect that our present studies would motivate experimentalists to synthesize the 2D monolayer Pt-d-BN material and other Pt-based single atom catalysts in the laboratory for future applications.

## Conflicts of Interest:

The authors have no additional conflicts of interest.

## AUTHOR INFORMATION

**Corresponding Author**
**Dr. Srimanta Pakhira** − *Theoretical Condensed Matter Physics and Advanced Computational Materials Science Laboratory, Department of Physics, Indian Institute of Technology Indore (IIT Indore), Simrol, Khandwa Road, Indore, Madhya Pradesh 453552, India.*

*Theoretical Condensed Matter Physics and Advanced Computational Materials Science Laboratory, Centre of Advanced Electronics (CAE), Indian Institute of Technology Indore, Indore, MP 453552, India.*
ORCID: orcid.org/0000-0002-2488-300X.
Email: spakhira@iiti.ac.in or spakhirafsu@gmail.com

**Authors**

**Mr. Lokesh Yadav** − *Theoretical Condensed Matter Physics and Advanced Computational Materials Science Laboratory, Department of Physics, Indian Institute of Technology Indore (IIT Indore), Simrol, Khandwa Road, Indore, Madhya Pradesh 453552, India.*

## Acknowledgement:

This work was financially supported by the Science and Engineering Research Board-Department of Science and Technology (SERB-DST), Government of India, under Grant No. CRG/2021/000572. Dr Srimanta Pakhira acknowledges the SERB-DST, Government of India, for providing his Early Career Research Award (ECRA) under project number ECR/2018/000255. Dr Pakhira also thanks the SERB-DST for providing the highly prestigious Ramanujan Faculty Fellowship under scheme number SB/S2/RJN-067/2017 and providing the highly prestigious Core Research Grant (CRG), SERB-DST, Govt. of India under the scheme number CRG/2021/000572. Ms. Lokesh thanks the CSIR, Govt. of India, and Govt. of India for providing his doctoral fellowship under scheme no. CSIRAWARD/JRF-NET2022/11898. The author would like to acknowledge the SERB-DST for providing computing clusters and programs. We thank the Council of Scientific and Industrial Research (CSIR), Government of India for providing the research grants under the scheme number 22/0883/23/EMR-II.

## Author Contributions:

Dr. Pakhira designed the project, and he conceived the complete idea of this current research project work, and Mr. Lokesh Yadav computationally studied the electronic structures and properties of the 2D monolayer h-BN, d-BN, and Pt-d-BN. Dr. Pakhira and Mr. Lokesh Yadav explored the whole reaction pathways, transition states, and reaction barriers. They explained the ORR mechanism by the DFT Quantum Mechanical calculations. Dr. Pakhira and Mr. Lokesh wrote the whole manuscript and prepared all the tables and figures in the manuscript. Dr. Pakhira and Mr. Lokesh interpreted and analyzed the computed results, and Dr. Pakhira supervised the project work.

**Graphical Abstract (TOC):**

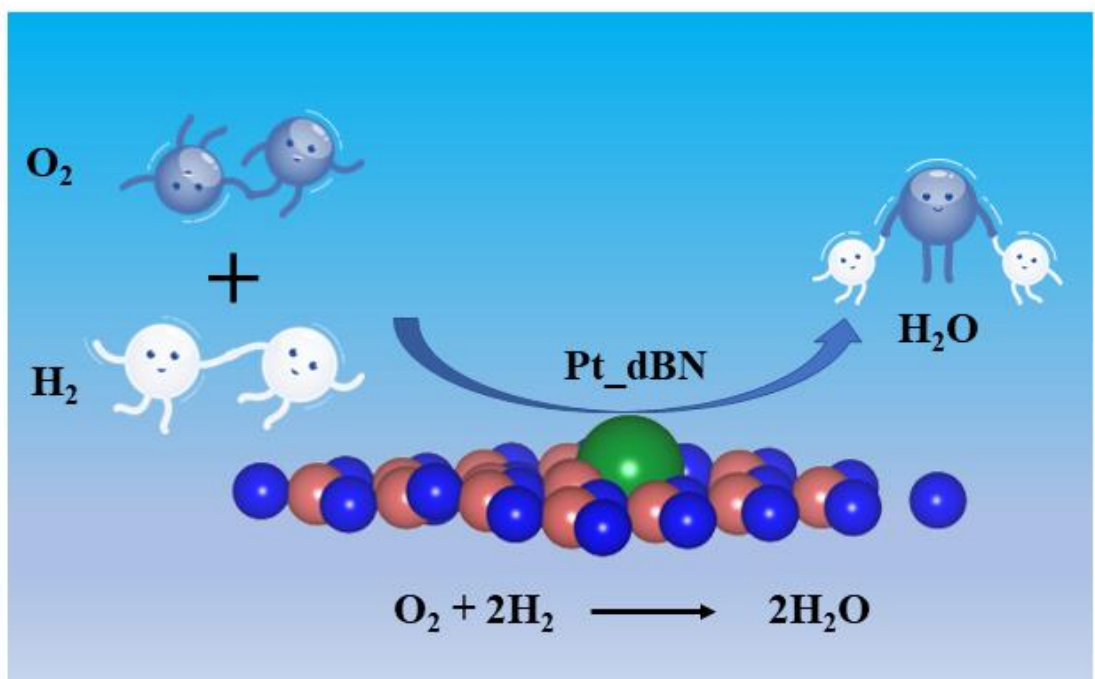